\documentclass[trackchanges,twocolumn]{aastex701}
\usepackage{lipsum}  
\usepackage[dvipsnames,table]{xcolor}
\usepackage{booktabs, multirow}
\usepackage[normalem]{ulem}
\usepackage{soul}
\usepackage{xspace}
\usepackage{lstautogobble} % For listings
\usepackage{placeins} % For \FloatBarrier
\usepackage{amsmath}

\usepackage{savesym}
\savesymbol{tablenum}
\usepackage{siunitx}
\restoresymbol{SIX}{tablenum}

\newcommand{\splus}{\mbox{S-PLUS}\xspace}
\newcommand{\ionn}[2]{{\textrm{#1}}{\textrm{\sc #2}}}
\newcommand{\hide}[1]{\ignorespaces}

\setcitestyle{notesep={; }}
\received{March 28, 2026}
\revised{May 6, 2026}
\submitjournal{ApJ}
\begin{document}

\title{The S-PLUS Fifth Data-Release: Over 4500 square degrees of the Southern Sky and a multicolor view of the Hydra and Antlia galaxy clusters}

\correspondingauthor{Erik Vinicius Rodrigues de Lima}
\email[show]{erikv.usp@gmail.com, erik.vini@alumni.usp.br}

\author[0000-0002-6268-8600]{Erik Vinicius Rodrigues de Lima}
\affiliation{Instituto de Astronomia, Geofísica e Ciências Atmosféricas, Universidade de São Paulo, Rua do Matão 1226, Cidade Universitária, São Paulo, SP, 05508-090, Brazil}
\email{erikv.usp@gmail.com}

% Gustavo:
\author[0009-0003-6609-1582]{Gustavo Bernhard Oliveira Schwarz}
\affiliation{Instituto de Astronomia, Geofísica e Ciências Atmosféricas, Universidade de São Paulo, Rua do Matão 1226, Cidade Universitária, São Paulo, SP, 05508-090, Brazil}
\email{gustavo.b.schwarz@gmail.com}
% Fábio: 
\author[0000-0001-7907-7884]{Fábio Rafael Herpich}
\affiliation{Laboratório Nacional de Astrofísica (LNA/MCTI), Rua Estados Unidos 154, Itajubá, MG, 37504-364, Brazil}
\email{fabiorafaelh@gmail.com}
% Felipe: 
\author[0000-0002-8048-8717]{Felipe Almeida-Fernandes}
\affiliation{Instituto Nacional de Pesquisas Espaciais, Av. dos Astronautas 1758, Jardim da Granja, São José dos Campos, SP, 12227-010, Brazil}
\email{felipe.fernandes@inpe.br}
%----- Lili: 
\author[0000-0001-6480-1155]{Lilianne Nakazono}
\affiliation{Observatório Nacional, Rua General José Cristino 77, São Cristóvão, Rio de Janeiro, RJ, 20921-400, Brazil}
\affiliation{Departamento de Física Matemática, Instituto de Física, Universidade de São Paulo, R. do Matão 1371, São Paulo, SP, 05508-090, Brazil}
\email{liliannenakazono@on.br}

%% Time de observação
% Eduardo Alberto Duarte Lacerda <
\author[0000-0001-7231-7953]{Eduardo Alberto Duarte Lacerda}
\affiliation{Instituto de Astronomia, Geofísica e Ciências Atmosféricas, Universidade de São Paulo, Rua do Matão 1226, Cidade Universitária, São Paulo, SP, 05508-090, Brazil}
\email{dhubax@gmail.com>}
%----- André Santos <>
\author[0000-0002-1420-3584]{André Santos}
\affiliation{Centro Brasileiro de Pesquisas Físicas, Rua Dr Xavier Sigaud 150, Rio de Janeiro, RJ, 22290-180, Brazil}
\email{andsouzasanttos@gmail.com}
%----- André: 
\author[0009-0007-1625-8937]{André Luiz Figueiredo}
\affiliation{Instituto de Astronomia, Geofísica e Ciências Atmosféricas, Universidade de São Paulo, Rua do Matão 1226, Cidade Universitária, São Paulo, SP, 05508-090, Brazil}
\email{figueiredo.iag@gmail.com}
% Luis Angel Gutiérrez Soto <>
\author[0000-0002-9891-8017]{Luis Angel Gutiérrez-Soto}
\affiliation{Instituto de Astrofísica de La Plata, UNLP-CONICET, Paseo del Bosque s/n, B1900FWA, La Plata, Argentina}
\affiliation{Instituto de Física, Universidade Federal do Rio Grande do Sul, Av. Bento Gonçalves 9500, Porto Alegre, RS, 90040-060, Brazil}
\email{gsoto.angel@gmail.com}
% Marília Jobim Sartori <>
\author[0000-0002-5931-5360]{Marília Jobim Sartori}
\affiliation{Instituto de Astronomia, Geofísica e Ciências Atmosféricas, Universidade de São Paulo, Rua do Matão 1226, Cidade Universitária, São Paulo, SP, 05508-090, Brazil}
\email{mariliajsartori@gmail.com}
% Marcos Antonio Fonseca Faria <>
\author[0000-0002-7865-3971]{Marcos Antonio Fonseca-Faria}
\affiliation{Laboratório Nacional de Astrofísica (LNA/MCTI), Rua Estados Unidos 154, Itajubá, MG, 37504-364, Brazil}
\email{mfaria@lna.br}
%----- Maiara: 
\author[0009-0008-9042-4478]{Maiara	Sampaio Carvalho}
\affiliation{Instituto de Astronomia, Geofísica e Ciências Atmosféricas, Universidade de São Paulo, Rua do Matão 1226, Cidade Universitária, São Paulo, SP, 05508-090, Brazil}
\email{mscarvalho.astro@gmail.com}
%----- "Bárbara Cubillos P." <>
\author[0009-0007-0020-0976]{Bárbara Cubillos Palma}
\affiliation{Departamento de Astronomía, Universidad de La Serena, Av. Raúl Bitrán 1305, La Serena, 1720256, Chile}
\email{bcubillospalma@gmail.com}

%% Escrita do paper
% Elismar: 
\author[0000-0003-2561-0756]{Elismar Lösch}
\affiliation{Instituto de Astronomia, Geofísica e Ciências Atmosféricas, Universidade de São Paulo, Rua do Matão 1226, Cidade Universitária, São Paulo, SP, 05508-090, Brazil}
\email{elismar.l@usp.br}
%----- Gissel: 
\author[0009-0003-1364-3590]{Gissel Pardo Montaguth}
\affiliation{Instituto de Astronomia, Geofísica e Ciências Atmosféricas, Universidade de São Paulo, Rua do Matão 1226, Cidade Universitária, São Paulo, SP, 05508-090, Brazil}
\email{gissel.pardo@userena.cl}
%Yuri: 
\author[0009-0002-2050-150X]{José Yuri Santos Silva}
\affiliation{Instituto de Astronomia, Geofísica e Ciências Atmosféricas, Universidade de São Paulo, Rua do Matão 1226, Cidade Universitária, São Paulo, SP, 05508-090, Brazil}
\email{yuricosmo@usp.br}
% Liana: 
\author[0000-0002-5295-7045]{Liana Li}
\affiliation{Instituto de Astronomia, Geofísica e Ciências Atmosféricas, Universidade de São Paulo, Rua do Matão 1226, Cidade Universitária, São Paulo, SP, 05508-090, Brazil}
\email{lianali1217@usp.br}
% Marina: 
\author[0009-0008-7546-5255]{Marina Izabela}
\affiliation{Instituto de Astronomia, Geofísica e Ciências Atmosféricas, Universidade de São Paulo, Rua do Matão 1226, Cidade Universitária, São Paulo, SP, 05508-090, Brazil}
\email{marina.izabela.mi@gmail.com}
%----- Natanael: 
\author[0000-0002-2238-9665]{Natanael Magalhães Cardoso} %\email{natanael.mc@usp.br}
\affiliation{Escola Politécnica, Universidade de São Paulo, São Paulo, SP, 05508-010, Brasil}
\email{natanael.mc@usp.br}
% Thais: 
\author[0000-0001-8789-6230]{Thais Santos-Silva}
\affiliation{Universidade Estadual de Feira de Santana, Av. Transnordestina s/n, Novo Horizonte, Feira de Santana, BA, 44036-900, Brazil}
\affiliation{Instituto de Astronomia, Geofísica e Ciências Atmosféricas, Universidade de São Paulo, Rua do Matão 1226, Cidade Universitária, São Paulo, SP, 05508-090, Brazil}
\email{thaisfi@gmail.com}

%% Pessoas que fizeram algo usado diretamente no paper
% Paulo Lopes: 
\author[0000-0003-2540-7424]{Paulo Afrânio Augusto Lopes}
\affiliation{Observatório do Valongo, Universidade Federal do Rio de Janeiro, Ladeira do Pedro Antônio 43, Rio de Janeiro, RJ, 20080-090, Brazil}
\email{plopes@ov.ufrj.br}
%----- Pedro Humire
\author[0000-0003-3537-4849]{Pedro K. Humire}
\affiliation{Instituto de Astronomia, Geofísica e Ciências Atmosféricas, Universidade de São Paulo, Rua do Matão 1226, Cidade Universitária, São Paulo, SP, 05508-090, Brazil}
\email{pedrohumirer@gmail.com}
% Roberta Vassallo: 
\author[0009-0002-0567-9561]{Roberta Vassallo Bordoni}
\affiliation{Instituto de Física, Universidade de São Paulo, Rua do Matão 1371, São Paulo, SP, 05508-090, Brazil}
\email{roberta.bordoni@usp.br}

%% Outros
% Amanda Lopes: 
\author[0000-0002-6164-5051]{Amanda Reis Lopes}
\affiliation{Instituto de Astronomia, Geofísica e Ciências Atmosféricas, Universidade de São Paulo, Rua do Matão 1226, Cidade Universitária, São Paulo, SP, 05508-090, Brazil}
\affiliation{Instituto de Astrofísica de La Plata, UNLP-CONICET, Paseo del Bosque s/n, B1900FWA, La Plata, Argentina}
\email{amandalopes@on.br}
% Analia: 
\author[0009-0007-2396-0003]{Analía Viviana Smith Castelli}
\affiliation{Instituto de Astrofísica de La Plata, UNLP-CONICET, Paseo del Bosque s/n, B1900FWA, La Plata, Argentina}
\affiliation{Facultad de Ciencias Astronómicas y Geofísicas, Universidad Nacional de La Plata, Paseo del Bosque s/n, B1900, La Plata, Argentina}
\email{a.smith.castelli@gmail.com}
% Angela Krabbe: 
\author[0000-0003-4630-1311]{Ângela Cristina Krabbe}
\affiliation{Instituto de Astronomia, Geofísica e Ciências Atmosféricas, Universidade de São Paulo, Rua do Matão 1226, Cidade Universitária, São Paulo, SP, 05508-090, Brazil}
\email{angela.krabbe@gmail.com}
%----- Augusto Damineli: 
\author[0000-0002-7978-2994]{Augusto Damineli}
\affiliation{Instituto de Astronomia, Geofísica e Ciências Atmosféricas, Universidade de São Paulo, Rua do Matão 1226, Cidade Universitária, São Paulo, SP, 05508-090, Brazil}
\email{augusto.damineli@iag.usp.br}
%----- Carlos Eduardo
\author[0000-0002-8525-7977]{Carlos Eduardo Ferreira Lopes}
\affiliation{Institute of Astronomy and Planetary Sciences, University of Atacama, Copayapu 485, Copiapó, Chile}
\email{ferreiralopes1011@gmail.com}
%----- Círia: 
\author[0009-0006-0373-8168]{Ciria Lima-Dias}
\affiliation{Departamento de Astronomía, Universidad de La Serena, Av. Raúl Bitrán 1305, La Serena, 1720256, Chile}
\email{clima@userena.cl}
% Clécio
\author[0000-0003-4383-2969]{Clécio Roque de Bom}
\affiliation{Centro Brasileiro de Pesquisas Físicas, Rua Dr Xavier Sigaud 150, Rio de Janeiro, RJ, 22290-180, Brazil}
\email{clecio@debom.com.br}
% Daniela Olave: 
\author[0000-0001-9243-3425]{Daniela E. Olave-Rojas}
\affiliation{Departamento de Tecnologías Industriales, Facultad de Ingeniería, Universidad de Talca, Los Niches km 1, Curicó, Chile}
\email{de.olaver@gmail.com}
%----- Debasish
\author[0000-0003-4379-6777]{Debasish Hazarika}
\affiliation{Institute of Astronomy and Planetary Sciences, University of Atacama, Copayapu 485, Copiapó, Chile}
\affiliation{NSF NOIRLab, Tucson, AZ 85719, USA}
\email{debasish.academic@gmail.com}
% Eduardo Telles
\author[0000-0002-8280-4445]{Eduardo Telles}
\affiliation{Observatório Nacional, Rua General José Cristino 77, São Cristóvão, Rio de Janeiro, RJ, 20921-400, Brazil}
\email{telles@on.br}
%----- Erick Ghuron
\author[0000-0002-2729-8869]{Erick Ghuron}
\affiliation{Instituto de Física, Universidade de São Paulo, Rua do Matão 1371, Cidade Universitária, São Paulo, SP, 05508-090, Brazil}
\email{ghuron@usp.br}
%----- Guilherme Limberg
\author[0000-0002-9269-8287]{Guilherme Limberg}
\affiliation{Kavli Institute for Cosmological Physics, University of Chicago, 5640 S. Ellis Avenue, Chicago, IL 60637, USA}
\affiliation{Department of Astronomy \& Astrophysics, University of Chicago, 5640 S. Ellis Avenue, Chicago, IL 60637, USA}
\email{limberg@uchicago.edu}
%----- Helio Perotonni
\author[0000-0002-0537-4146]{Hélio Dotto Perottoni}
\affil{Observatório Nacional, Rua General José Cristino 77, São Cristóvão, Rio de Janeiro, RJ, 20921-400, Brazil}
\email{hperottoni@gmail.com}
% Júlia Thainá: 
\author[0009-0008-2216-9575]{Júlia Thainá-Batista}
\affiliation{Instituto de Astronomia, Geofísica e Ciências Atmosféricas, Universidade de São Paulo, Rua do Matão 1226, Cidade Universitária, São Paulo, SP, 05508-090, Brazil}
\email{jullia.thainna@gmail.com}
%----- Laerte: 
\author[0000-0002-3876-268X]{Laerte Sodré Jr.}
\affiliation{Instituto de Astronomia, Geofísica e Ciências Atmosféricas, Universidade de São Paulo, Rua do Matão 1226, Cidade Universitária, São Paulo, SP, 05508-090, Brazil}
\email{laerte.sodre@iag.usp.br}
%----- Lia: 
\author[0000-0001-8450-5193]{Lia Doubrawa}
\affiliation{Department of Physics, University of Helsinki, P.O. Box 64, FI-00014 Helsinki, Finland}
\affiliation{Instituto de Astronomia, Geofísica e Ciências Atmosféricas, Universidade de São Paulo, Rua do Matão 1226, Cidade Universitária, São Paulo, SP, 05508-090, Brazil}
\email{lia.doubrawa@usp.br}
%----- Marcelo Borges: 
\author[0000-0001-5740-2914]{Marcelo Borges Fernandes}
\affiliation{Observatório Nacional, Rua General José Cristino 77, São Cristóvão, Rio de Janeiro, RJ, 20921-400, Brazil}
\email{borges@on.br}
% Murillo
\author[0000-0001-9719-4523]{Murillo Marinello}
\affiliation{Laboratório Nacional de Astrofísica (LNA/MCTI), Rua Estados Unidos 154, Itajubá, MG, 37504-364, Brazil}
\email{mmarinello@lna.br}
%----- Panda
\author[0000-0002-5854-7426]{Swayamtrupta Panda} \thanks{Gemini Science Fellow}
\affiliation{International Gemini Observatory/NSF NOIRLab, Casilla 603, La Serena, Chile}
\email{swayamtrupta.panda@noirlab.edu}
%----- Pierre: 
\author[0000-0002-2008-8615]{Pierre Augusto Ré}
\affiliation{Instituto de Astronomia, Geofísica e Ciências Atmosféricas, Universidade de São Paulo, Rua do Matão 1226, Cidade Universitária, São Paulo, SP, 05508-090, Brazil}
\email{pierre@usp.br}
%----- Raimundo Lopes Oliveira filho: 
\author[0000-0002-6211-7226]{Raimundo Lopes de Oliveira}
\affiliation{Departamento de Física, Universidade Federal de Sergipe, Av. Marechal Rondon s/n, São Cristóvão, SE, 88040-900, Brazil}
\affiliation{Observatório Nacional, Rua General José Cristino 77, São Cristóvão, Rio de Janeiro, RJ, 20921-400, Brazil}
\email{raimundo.lopes@academico.ufs.br}
%----- Raquel
\author[0000-0001-8847-0047]{Raquel Ruiz Valença}
\affiliation{Instituto de Astronomia, Geofísica e Ciências Atmosféricas, Universidade de São Paulo, Rua do Matão 1226, Cidade Universitária, São Paulo, SP, 05508-090, Brazil}
\email{vruizraquel@usp.br}
% Ricardo DeMarco: 
\author[0000-0003-3921-2177]{Ricardo Demarco}
\affiliation{Institute of Astrophysics, Facultad de Ciencias Exactas, Universidad Andrés Bello, Sede Concepción, 4260000, Talcahuano, Chile}
\email{rdemarco@astro-udec.cl}
% Roberto Cid: 
\author[0000-0001-9672-0296]{Roberto Cid-Fernandes}
\affiliation{Departamento de Física, Universidade Federal de Santa Catarina, Florianópolis, SC, 88040-900, Brazil}
\email{robertocidfernandes@gmail.com}
% Rodrigo Haack: 
\author[0009-0005-6830-1832 ]{Rodrigo Facundo Haack}
\affiliation{Instituto de Astrofísica de La Plata, UNLP-CONICET, Paseo del Bosque s/n, B1900FWA, La Plata, Argentina}
\affiliation{Facultad de Ciencias Astronómicas y Geofísicas, Universidad Nacional de La Plata, Paseo del Bosque s/n, B1900, La Plata, Argentina}
\email{rodrihaack@gmail.com}
% Sergio Torres-Flores: 
\author[0000-0002-7005-8983]{Sergio Torres-Flores}
\affiliation{Departamento de Astronomía, Universidad de La Serena, Av. Raúl Bitrán 1305, La Serena, 1720256, Chile}
\email{spatorres@gmail.com}
%----- Stavros Akras
\author[0000-0003-1351-7204]{Stavros Akras}
\affiliation{Institute for Astronomy, Astrophysics, Space Applications and Remote Sensing, National Observatory of Athens, GR 15236 Penteli, Greece}
\email{stavrosakras@gmail.com}
%----- Tim Beers
\author[0000-0003-4573-6233]{Timothy C. Beers}
\affiliation{Department of Physics and Astronomy, University of Notre Dame, Notre Dame, IN 46556, USA}
\affiliation{Joint Institute for Nuclear Astrophysics, Center for the Evolution of the Elements (JINA-CEE), East Lansing, MI 48824, USA}
\email{Timothy.C.Beers.5@nd.edu}
%----- Vinicius Placco
\author[0000-0003-4479-1265]{Vinicius M. Placco}
\affiliation{NSF NOIRLab, Tucson, AZ 85719, USA}
\email{vmplacco@gmail.com}
%----- Vitor Cernic
\author[0000-0002-1233-0301]{Vitor Cernic}
\affiliation{Independent Researcher}
\email{vitorcernic@gmail.com}
% Victor Hugo Sasse
\author[0009-0008-8763-7050]{Victor Hugo Sasse}
\affiliation{Universidade Federal de Santa Catarina, Campus Universitário Reitor João David Ferreira Lima, Florianópolis, SC, 88040-900, Brazil }
\email{victorsasse62@gmail.com}

%% Builders
% A. Kanaan
\author[0009-0007-8005-4541]{Antonio Kanaan}
\affiliation{Departamento de Física, Universidade Federal de Santa Catarina, Florianópolis, SC, 88040-900, Brazil}
\email{ankanaan@gmail.com}
% T. Ribeiro
\author[0000-0002-0138-1365]{Tiago Ribeiro}
\affiliation{Rubin Observatory Project Office, 950 N. Cherry Ave, Tucson 85719, USA}
\email{tiago.astro@gmail.com}
% W. Schoennell
\author[0000-0002-4064-7234]{William Schoennell}
\affiliation{The Observatories of the Carnegie Institution for Science, 813 Santa Barbara St, Pasadena, CA 91101, USA}
\email{wschoenell@gmail.com}
% Cláudia
\author[0000-0002-5267-9065]{Cláudia Lúcia Mendes de Oliveira}
\affiliation{Instituto de Astronomia, Geofísica e Ciências Atmosféricas, Universidade de São Paulo, Rua do Matão 1226, Cidade Universitária, São Paulo, SP, 05508-090, Brazil}
\email{claudia.oliveira@iag.usp.br}

%\author{the S-PLUS Collaboration}

% Reset footnote counter
\let\oldmaketitle\maketitle
\renewcommand{\maketitle}{\oldmaketitle\setcounter{footnote}{0}}
\collaboration{all}{(S-PLUS Collaboration)}

%% Use the \collaboration command to identify collaborations. This command
%% takes an optional argument that is either a number or the word "all"
%% which tells the compiler how many of the authors above the command to
%% show. For example "\collaboration[all]{(DELVE Collaboration)}" wil include
%% all the authors above this command.
%%
%% Mark off the abstract in the ``abstract'' environment. 
\begin{abstract}

We present the 5th data release (DR5) of the Southern Photometric Local Universe Survey (\splus), covering 4592 square degrees across 2491 fields. Observations were conducted with the T80-South, a Brazilian robotic telescope equipped with the Javalambre 12-filter system, containing five broad- and seven narrowband filters. Data products feature FITS images and extensive catalogs containing fluxes, magnitudes, and shape parameters for over 113 million detections. In addition, several value-added catalogs are provided, offering photometric redshifts (photo-zs), object classifications, masks, and extinction coefficients. For the first time, this release includes coverage of 110 square degrees along the Galactic disk, facilitating new research into Galactic structure and stellar populations. The release also provides full coverage of the Hydra Supercluster and numerous other nearby clusters with improved data reduction and calibration, enhancing photo-z accuracy, which is vital for large-scale structure studies. A preliminary analysis of the Hydra and Antlia galaxy clusters up to $5 \times R_{200}$ yields an updated catalog of 1706 cluster members based on both spectroscopic and high-quality photo-zs. Our photometric data shows that both clusters have a similar proportion of galaxies with an H$\alpha$ excess relative to their clustercentric distance, though Hydra has a higher fraction near its center. Additionally, the spatial distribution of all objects in our sample highlights a bridge connecting both clusters. We verify that \splus DR5 provides a solid foundation for future scientific investigations, ranging from Solar System studies to Cosmology.

\end{abstract}

%% Keywords should appear after the \end{abstract} command. 
%% The AAS Journals now uses Unified Astronomy Thesaurus concepts:
%% https://astrothesaurus.org
%% You will be asked to select these concepts during the submission process
%% but this old "keyword" functionality is maintained in case authors want
%% to include these concepts in their preprints.
%\keywords{S-PLUS -- Catalogs -- Galaxy: general -- Stars: general -- Galaxies: general}

%% From the front matter, we move on to the body of the paper.
%% Sections are demarcated by \section and \subsection, respectively.
%% Observe the use of the LaTeX \label
%% command after the \subsection to give a symbolic KEY to the
%% subsection for cross-referencing in a \ref command.
%% You can use LaTeX's \ref and \label commands to keep track of
%% cross-references to sections, equations, tables, and figures.
%% That way, if you change the order of any elements, LaTeX will
%% automatically renumber them.

% Introduction
\section{Introduction} \label{sec:intro}

In recent years, astronomers have increasingly relied on archival and survey data rather than individual targeted observations. In this context, wide-field surveys are fundamental tools in astronomy, as they enable efficient observations over extended areas of the sky (e.g the SkyMapper survey, \citealt{Onken2024-SkyMapper}). Some multiband photometric surveys also provide a detailed description of the spectral energy distribution of objects, especially when utilizing narrowbands. This allows the determination of the physical properties of galaxies, such as their stellar masses, star formation rates, and stellar population parameters \citep{2016ApJS..227....2S,2017MNRAS.471.4722M,2023MNRAS.526.1874T,Humire2025}. Additionally, they enable the calculation of high-precision photometric redshifts \citep{2000A&A...363..476B,2022A&C....3800510L} for cosmic evolution studies, as well as the detection of transient events through temporal variability.

\begin{figure*}[ht!]
    \includegraphics[width=\linewidth]{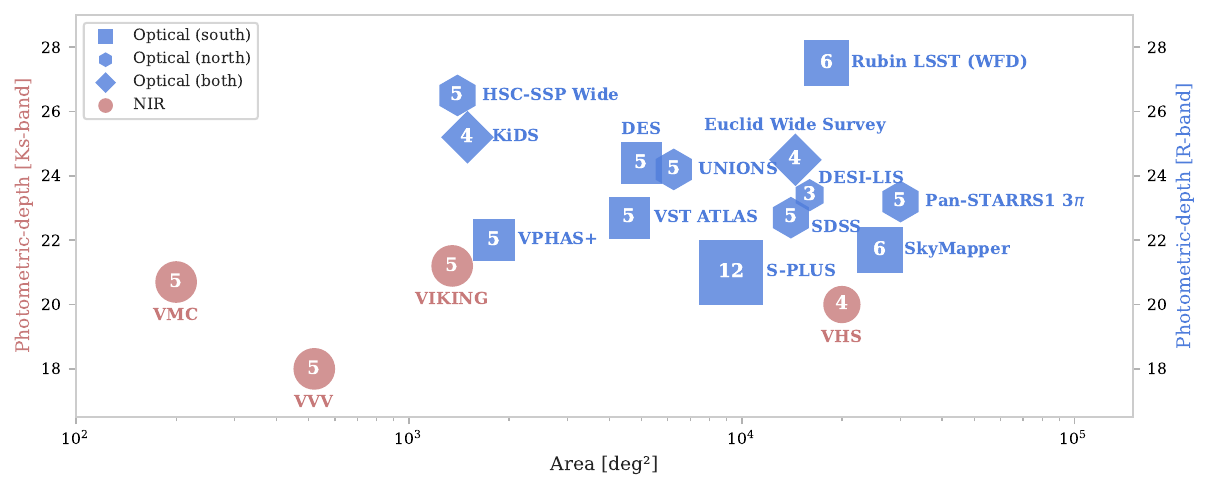}
    \caption{Comparison of several multiband surveys for both hemispheres in the optical (blue shapes) and NIR (red circles). The y-axis on either side indicate their depth in AB magnitudes. The number inside the shapes represent the number of photometric filters of the survey.}
    \label{fig:survey_comparison}
\end{figure*}

Multiband photometric surveys have also been extensively employed to determine stellar atmospheric parameters \citep{Kunzli+1997, Casagrande+2010} and abundances \citep{Ivezic+2008}. These provide important constraints for studies of our Galaxy as a whole, such as the identification and characterization of satellite galaxies \citep{Drlica-Wagner+2015,Cerny2025,Tan2025}, stellar streams \citep{Belokurov+2006,Shipp2018}, and the overall structure of the Milky Way \citep{Juric+2008,Limberg2025,Placco2025}. They also have enabled the identification of rare and peculiar objects that deserve targeted spectroscopic follow-ups, such as the stars in the lowest metallicity regime \citep{Starkenburg+2017} and point sources with high H$\alpha$ emission \citep{Witham+2008,Soto2025}. A comparison between multiband photometric surveys is presented in Figure \ref{fig:survey_comparison}.

Among those projects, the Southern Photometric Local Universe Survey \citep[\splus;][]{2019MNRAS.489..241M} stands out as the one with the highest number of filters dedicated to observing the Southern Hemisphere. It uses the Javalambre 12-filter photometric system \citep{2019A&A...622A.176C}, which consists of the Javalambre $u_\text{JAVA}$ (referred to as $u$ throughout this work), four SDSS-like bands ($g$, $r$, $i$, and $z$), and seven narrowbands. The narrowbands $J0378$, $J0395$, $J0410$, $J0430$, $J0515$, $J0660$, and $J0861$ are specifically designed to capture key spectral features, allowing for detailed studies of stellar populations and ionized gas \citep{navarro2012modern}. In particular, \splus aims to cover an area of \num{9300} square degrees on the Southern sky using a dedicated 0.8m robotic telescope, the T80-South, located at the NSF Cerro Tololo Inter-American Observatory in Chile (see \citealt{2019MNRAS.489..241M} for more details). In this manuscript, we present the fifth data release (hereafter DR5), which expands the coverage to a total of $\sim$\num{4600} square degrees on its Main Survey.

\splus narrowband photometry has already proven to be reliable for stellar parameter estimation through machine learning (ML) techniques \citep{Whitten+2021, QuispeHuaynasi+2024, MolinaJorquera+2024, FerreiraLopes+2025}. This enables stellar population studies, such as metallicity trends, kinematics, and spatial distributions in the Sagittarius Stream and halo substructures (Almeida-Fernandes et al. in prep.; Bolutavicius et al. in prep.; \citealt{Johnson2020,Cunningham+2024,Helmi2020,Deason2024}), providing constraints on their origins and assembly histories. \splus has also enabled searches for ultra-metal-poor stars \citep{Placco+2022, AlmeidaFernandes+2023} and the identification of H$\alpha$ sources (e.g., interacting binaries/YSOs/nebulae and AGN; \citealt{Soto2025}). Notably, the metal-poor searches have yielded record-setting objects, including a star with the lowest measured carbon abundance for its type at the time \citep{Placco+2021} and an extremely metal-poor star with the highest Th/Eu ratio observed to date \citep{Placco2023}. Additionally, the \splus DR4 coverage of the Magellanic Cloud region enables studies of variable stars, such as Cepheids and RR Lyrae stars, for both fundamental and first-overtone pulsation modes. This sample has been used to derive 12-band, single-epoch Cepheid period-luminosity relations (PLR) and to construct Cepheid-based 3D reddening maps of the Small and Large Magellanic Clouds (Hazarika D. et al., submitted to A\&A).

DR5 is the first data release to cover the area toward the bulge of the Milky Way, allowing the targeting of additional low-metallicity stars for spectroscopic follow-ups. Combined with the Visible and Infrared Survey Telescope for Astronomy (VISTA) Variables in the Via Lactea survey \citep[VVV;][]{Minniti+2010} and the upcoming high-multiplex spectrograph 4-metre Multi-Object Spectroscopic Telescope \citep[4MOST;][]{deJong+2019} of the European Southern Observatory (ESO), DR5 provides valuable data to explore this region. We also note that \splus contains dedicated sub-surveys, such as the Ultra-Short Survey, designed to optimize the identification of bright metal-poor stars \citep{Perottoni2024}. Moreover, DR5 extends variable-star work across the bulge and disk, with added information from narrowbands. In addition, DR5 can test the Cepheid PLR-[Fe/H] dependence, which is relevant for stellar pulsation models and cosmic distance-scale systematics. This also informs distance-ladder versus CMB comparisons, although PLR metallicity effects alone have been shown to be insufficient to reconcile the Hubble tension \citep{2022Riess}. Because spectroscopic [Fe/H] is limited for large Galactic Cepheid samples, \splus metallicity-sensitive bands plus the DR5 bulge and disk coverage provide a strong alternative.

As a wide-field survey, \splus enables the identification of new galaxy groups and clusters \citep{10.1093/mnras/stac3273}, allowing studies of both pre- and post-processing (\citealt{2014A&A...570A.119E,2015ApJ...806..101H,2023MNRAS.524.5340M,2024MNRAS.528.4993C}). In these evolutionary stages, galaxies experience environmental effects in groups or filaments prior to or after entering the denser environment, and they may continue to evolve while remaining coherent and resisting disruptive tidal forces or external perturbations for up to one orbital period \citep{2023ApJ...954...98P, 2021MNRAS.500.1323L, 2024MNRAS.528.4993C,2023MNRAS.519.4171O, 2024MNRAS.530.3787S, Haack2026}. By analyzing galaxies across different large-scale structures, from small groups to massive clusters, DR5 enables the investigation of environmental effects such as tidal interactions \citep{1996Natur.379..613M,1972AJ.....77..288G}, ram-pressure stripping \citep[RPS;][]{1972ApJ...176....1G,2015MNRAS.448.1715J}, and mergers (\citealt{1972ApJ...178..623T}; \citealt{Kaviraj2025-Mergers}), offering a comprehensive view of the physical processes driving galaxy transformation and helping us to unravel the complex role of the local environment in shaping galaxy properties over time (\citealt{Gissel2026a}; \citeyear{Gissel2026b}).

\splus DR5 photometry enables galaxy-cluster membership assignments out to $5\times R_{200}$\footnote{$R_{200}$ is defined as the radius in which the density of the cluster is 200 times the critical density of the Universe at its redshift.} of galaxy clusters (Lösch et al., in prep). This capability can be applied to the survey's photometric data to characterize both cluster cores and infall regions. For instance, \citet{Lima-Dias+2024} used \splus for multiwavelength bulge-disk decompositions in Hydra, linking environment to star-formation suppression. Additionally, tools like \mbox{\texttt{ASTROMORPHLIB}} can be used to derive nonparametric morphologies (e.g., CAS, Gini) to identify ram-pressure stripping candidates \citep{Hernandez-Jimenez&Krabbe2022, Krabbe+2024}. With DR5, these and related studies can be expanded to larger cluster radii and include additional cluster systems in the \splus footprint. 
Furthermore, \splus has proven to be crucial for determining cluster membership in the CHileAN Cluster galaxy Evolution Survey (CHANCES) and \splus Clusters and their Large-scale Environments (SCALE) projects (\citealt{chances}; Mendes de Oliveira et al., submitted.), especially in their outskirts, where contamination from background galaxies is significant (\citealt{Sifon2025}; \citealt{MendezHernandez2025}; Dobrawa et al. 2026, accepted).
%Furthermore, \splus has proven to be crucial for determining cluster membership in the CHileAN Cluster galaxy Evolution Survey \citep[CHANCES,]{chances} and \splus Clusters and their Large-scale Environments (SCALE) projects (\citealt{Sifon2025}; Mendes de Oliveira et al., in prep.), especially in their outskirts, where contamination from background galaxies is significant \citep{MendezHernandez2025,Doubrawa2025}.

In this context, a key DR5 highlight is the first homogeneous imaging of the entire Antlia cluster \citep{1989BAAS...21..780H,2012MNRAS.419.2472S,Calderon2020}, alongside the Hydra cluster \citep{2008A&A...486..697M,2011ApJS..197...31S}, presented in DR3. The Hydra-Antlia region at redshift $z \sim 0.01$ is a large-scale structure in the nearby Universe \citep{1995A&A...297..617K}. Despite its proximity to the Zone of Avoidance, several studies have explored the extent of a filament, which is sometimes referred to as the Hydra Wall \citep{2024MNRAS.531.3486R}. Therefore, the DR5 dataset provides valuable data on the Hydra supercluster \citep{2000A&AS..141..123K}, as it covers an area spanning from $-20$ to $-48$ degrees in declination and from 10h to 11h in right ascension, enabling a comprehensive study of the substructures and filaments of this region. Moreover, the broader coverage of \splus in the Hydra-Antlia region was also used to search for jellyfish galaxy candidates, as demonstrated by \citet{2024MNRAS.532..270G}.

Now in its ninth year, \splus has observed $\sim$50\% of its planned footprint. This large dataset enables extensive synergies with ongoing and upcoming wide-area surveys (e.g., Gaia, SDSS-V, DESI, Euclid, LSST, Roman; \citealt{Gaia2016, Almeida2023-SDSSV, DESI2022, Euclid2022, LSST2019, Spergel2015-WFIRST}). Substantial footprint overlap with 10- and 30-meter class facilities will further enhance these synergies. \splus multi-year observations also complement other time-domain surveys (e.g., ZTF; \citealt{Belim2014-ZTF}). Additionally, \splus Target of Opportunity and Variability (ToO/Var) programs provide high-cadence (almost daily), multiband monitoring. For example, monitoring of the AGN III Zw 02 combined with SOAR/GEMINI spectroscopy has recently revealed variations in its accretion disk and broad-line region (\citealt{DiasDosSantos2023}; Panda et al., in prep.). \splus thus holds promising prospects in the upcoming years with similar programs.

This paper is structured as follows: Section \ref{sec:obs_dataprocess} summarizes the observations and data reduction, including the new pipeline and Gaia-based calibration. Section \ref{sec:data} characterizes the data in terms of photometric depth and completeness, while Section \ref{sec:cat} describes the DR5 catalogs and their value-added products. Finally, Section \ref{sec:science} presents a science application to the Hydra-Antlia region, and Section \ref{sec:conclusion} provides the conclusions.

% Observations and data processing
\section{Observations and Data Processing} 
\label{sec:obs_dataprocess}

\subsection{What Is New in \splus DR5} \label{sec:obs-summary}

The fifth data release of the Southern Photometric Local Universe Survey (\splus DR5) represents a substantial expansion and consolidation of the survey relative to the previous public release \citep[DR4;][]{2024A&A...689A.249H}. The most significant developments are summarized below, with more detailed explanations outlined in the following sections.

\begin{itemize}
    \item Expanded sky coverage: DR5 increases the surveyed area from approximately 3000 to $\sim$4600 square degrees, corresponding to 2491 observed fields. This expansion includes new regions at low Galactic latitude and provides, for the first time, systematic coverage of fields adjacent to the Galactic bulge, substantially broadening the survey's Galactic science reach.
    
    \item Updated data reduction pipeline: A new data reduction framework, the Multiband Astronomical Reduction (MAR) pipeline, is employed for the majority of DR5 fields. MAR delivers improved astrometric solutions in crowded stellar environments, enabling reliable processing of dense disk and bulge-adjacent regions that were not accessible in earlier releases. Observations obtained prior to 2018 are retained with the previous pipeline to preserve homogeneity.
    
    \item Refined photometric products: DR5 builds on the photometric framework introduced in DR4 by further refining the balance between aperture and PSF photometry on a field-by-field basis. In addition, a new class of restricted AUTO magnitudes is introduced, designed to improve signal-to-noise while maintaining sensitivity to the total flux of compact and faint sources.
    
    \item Revised photometric calibration: A major methodological change in DR5 is the adoption of a Gaia DR3–spectra–based photometric calibration. Synthetic magnitudes derived from Gaia spectra are used as calibration references, yielding a uniform photometric scale across the entire footprint and naturally accounting for interstellar extinction. This approach removes the need for region-dependent corrections and leads to improved consistency in photometric measurements.
    
    \item Enhanced value-added catalogs: The improvements in calibration and depth translate directly into more accurate value-added products, including photometric redshifts, star–galaxy–quasar classifications, and extinction-related quantities. These enhancements are particularly relevant for studies of large-scale structure and galaxy environments.
    
    \item Extended cluster and large-scale structure science: DR5 provides homogeneous coverage of the Hydra–Antlia region, enabling cluster membership determination out to $5\times R_{200}$ and facilitating studies of filaments and inter-cluster connections. This represents a significant advance beyond the cluster-scale demonstrations presented in DR4.
\end{itemize}

Overall, DR5 constitutes a significant advancement in both survey area and data uniformity, establishing a more comprehensive and robust foundation for Galactic, extragalactic, and large-scale structure studies using \splus data.

\subsection{Observations} \label{sec:obs-observations}

The \splus observations are taken with the T80-South telescope, an 86 cm Ritchey-Chrétien Cassegrain telescope located at the NSF Cerro Tololo Interamerican Observatory in Chile. The detector is an e2v $9232\times9216\ 10\,\mu m$-pixel CCD with a $1.4\times1.4$ square degree field-of-view and a 0.55\arcsec~plate scale at the focal plane of the telescope, with a typical seeing of 1.5 to 2\arcsec. The readout uses 16 amplifiers placed in an array of $8\times2$.

\begin{figure*}[ht!]
    \includegraphics[width=\linewidth]{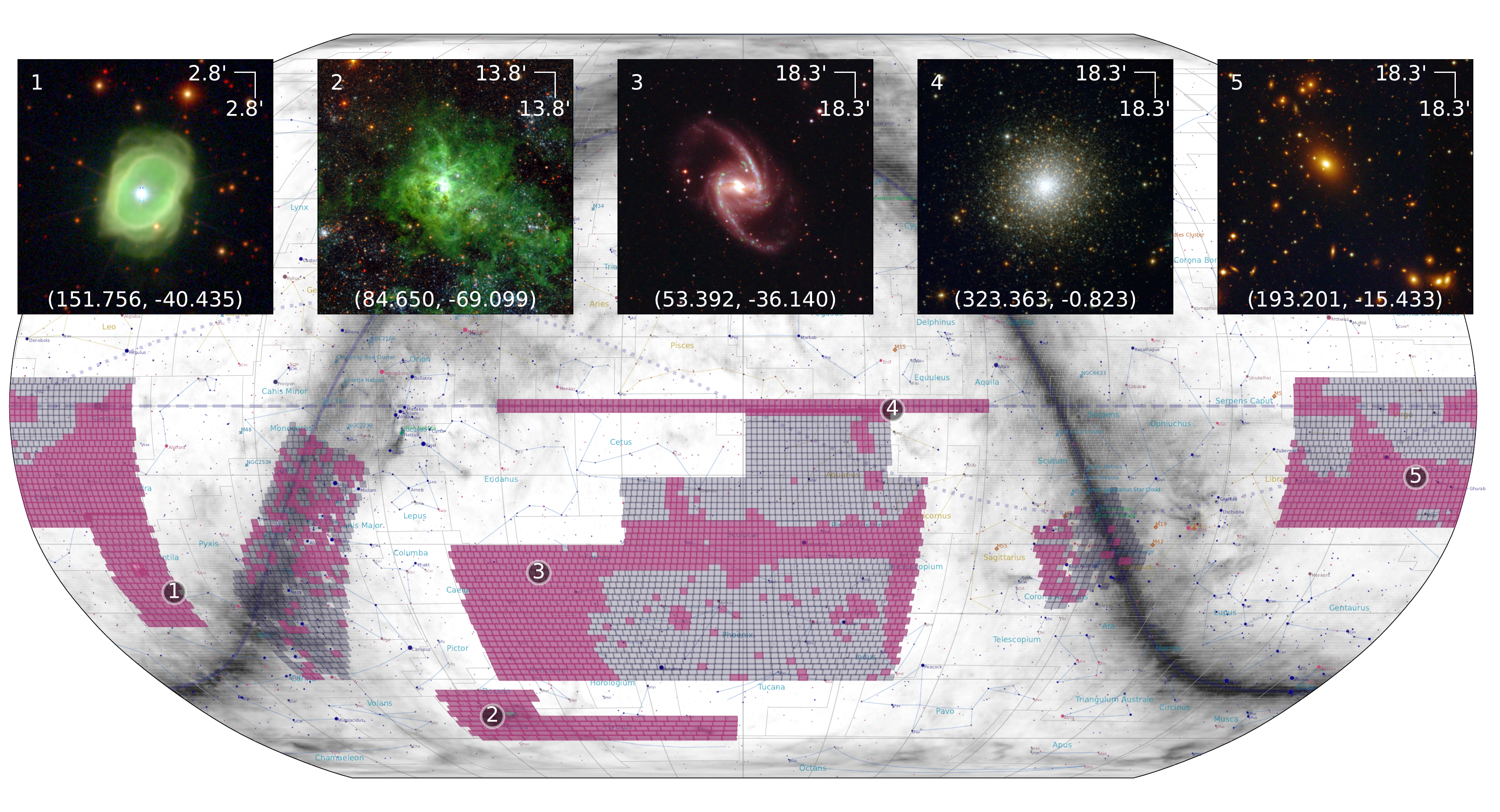}
    \caption{The footprint of \splus with squares representing the tiles. Grey squares represent those not yet observed, while the fuchsia ones represent the tiles available in DR5. The insets show examples of \splus images of different objects as indicated on the map. From left to right: The Southern Ring Nebula (NGC 3132), the 30 Doradus Region, the NGC 1365 galaxy in the Fornax Cluster, the M2 globular cluster, and the Abell 1631 cluster.}
    \label{fig:footprint}
\end{figure*}

For DR5, the observations were taken in a six-year campaign from August 2016 to July 2022 with two significant gaps: between April and November 2017, due to a series of technical issues, and between March and November 2020 due to COVID-19 restrictions. All observations undergo quality control on a nightly basis, via inspection of all images produced, to achieve an average seeing per field between 0.8 and 2.2\arcsec~during dark/gray photometric time for the Main Survey and bright time for the Galactic Survey (MS and GS, respectively, as defined in \citealt{2019MNRAS.489..241M}). The standard observation strategy involves observing the same field three times per filter with the aim of reaching depths of 21.5 mag in the $r$ band, using the exposure times listed in Table 5 of \citet{2019MNRAS.489..241M}. The DR5 coverage ($\sim4600$ square degrees in the Southern Hemisphere) corresponds to a total of 2491 tiles, which are shown in fuchsia in Figure \ref{fig:footprint}\footnote{The colored images in Figure \ref{fig:footprint} were constructed using the following stacking of filters for the (R, G, B) channels: NGC 3132 and the M2 globular cluster: (\{$r$, $i$, $J0861$, $z$\}, \{$g$, $J0515$, $J0660$\}, \{$u$, $J0378$, $J0395$, $J0410$, $J0430$\}); 30 Doradus: (\{$r$, $g$, $J0861$, $i$\}, \{$J0660$\}, \{$u$, $J0378$, $J0395$, $J0410$, $J0430$, $J0515$\}); NGC 1365: (\{$r$, $i$, $z$\}, \{$J0515$, $J0660$\}, \{$u$, $J0430$, $g$\}); Abell 1631 cluster: (\{r, $i$, $z$\}, \{$g$, $J0660$\}, \{$u$, $J0430$, $J0410$\})}.

\subsection{Data reduction} \label{sec:obs-reduction}

For observations obtained after January 1, 2018, we used the Multiband Astronomical Reduction pipeline \citep[MAR;][]{SCHWARZ2025100899}, corresponding to 2200 fields in DR5. The remaining 291 fields included in this release were reduced with the legacy \texttt{jype} pipeline, as in previous data releases \citep[see][]{2019MNRAS.489..241M,2022MNRAS.511.4590A,2024A&A...689A.249H}. These fields were kept in the legacy reduction stream because they were observed before 2018 and require detector non-linearity correction procedures that are not implemented in MAR. The two reduction streams are largely equivalent in their final calibrated products and provide consistent photometric and astrometric results for survey applications. MAR performs the standard reduction of raw astronomical images (see \citealt{SCHWARZ2025100899} for a full description), and its main practical improvement relative to the previous software is a more robust astrometric solution in crowded stellar fields, which enabled the inclusion of Galactic bulge and disk fields in DR5 for the first time.

The image reduction follows the sequence of overscan correction, trimming, bias subtraction, flat-field division, artifact corrections (including the removal of satellite trails and cosmic rays\footnote{We note that, in \splus DR5, an issue related to the reduction resulted in deficient cosmic-ray removal. However, we provide a Python-based method to detect and flag contaminated objects. This script is available in the DR5 documentation.}), and fringing correction in the $z$ band. No separate dark-current subtraction is applied, since the dark-current rate of the T80Cam e2v CCD is negligible for the exposure times and operating conditions of S-PLUS \citep{2019MNRAS.489..241M}. Dithered exposures are acquired in each filter for every pointing and are combined to produce the final publicly available images. During image combination, the resampling/coaddition procedure generates weight maps that provide pixel-level uncertainty estimates. These weight maps are used in subsequent analysis steps, such as photometric measurements and flux calibration, to propagate uncertainties.

\subsubsection{Obtaining the photometry} \label{sec:obs-photometry}

\splus DR5 provides point-spread function (PSF) photometry for 371 fields and aperture photometry for 2120 fields. When aperture photometry is available, the \texttt{PSTOTAL} aperture is the preferred choice for measuring the total magnitude of point sources. Several other options are also provided. These include adaptive apertures (\texttt{AUTO}, \texttt{PETRO}, \texttt{ISO}), which account for the shape and angular extent of sources, making them more suitable for extended objects, as well as fixed apertures (\texttt{APER\_3}, \texttt{APER\_6}, \texttt{PSTOTAL}). These apertures are defined in \citet{2024A&A...689A.249H}. Additionally, DR5 introduces restricted auto magnitudes (\texttt{RES}), which, like \texttt{AUTO}, are based on the Kron radius but use more constrained \texttt{Source-Extractor} \citep{Bertin1996-SExtractor} parameters\footnote{The restricted auto magnitudes are obtained by running \texttt{Source-Extractor} with \texttt{PHOT\_AUTOPARAMS} = 1.0, 1.0}. The restricted auto aperture is designed to improve signal-to-noise (S/N) ratio, while also taking into account the extended shape of the source. %at the cost of not capturing the total flux of the source.

As in previous data releases, and further described in \citet{2019MNRAS.489..241M}, aperture photometry is available in single and dual modes. In dual-mode photometry, all detections are made in a $griz$ co-added image. In single-mode photometry, detections and measurements are independent for each filter and are subsequently cross-matched with a 5\arcsec~ tolerance. This tolerance is much larger than astrometric errors ($\approx0.1$ arcsec, \citealt{2024A&A...689A.249H}) and was chosen to accommodate extended sources which can have different centroids depending on the filter of the observation. For studies of emission lines detected in narrowband filters, single-mode aperture photometry is expected to yield the most reliable detections. We note this 5\arcsec~tolerance increases the incidence of ambiguous/spurious matches for compact sources in crowded regions, so dual-mode photometry is generally preferable in these cases.

\subsubsection{Photometric calibration} \label{sec:obs-phot_calibration}

An accurate photometric calibration is essential for deriving reliable stellar and extragalactic properties, including photometric redshifts and stellar metallicities.

For \splus DR5, we implemented a new photometric calibration methodology based on Gaia DR3 spectra \citep{GaiaDR3-2023,Angeli2023}. This methodology offers two key advantages over previous implementations: First, it provides a uniform reference for photometric calibration across the entire \splus footprint without the need for \textit{ad hoc} corrections. Second, because we use the spectra obtained for the same stars that are being observed, this methodology naturally accounts for interstellar extinction in the reference magnitudes, eliminating the need for extinction corrections in the initial calibration step. These improvements significantly enhance the precision and reliability of \splus photometry, benefiting a wide range of astrophysical applications.

\begin{figure*}
    \centering
    \includegraphics[width=\linewidth]{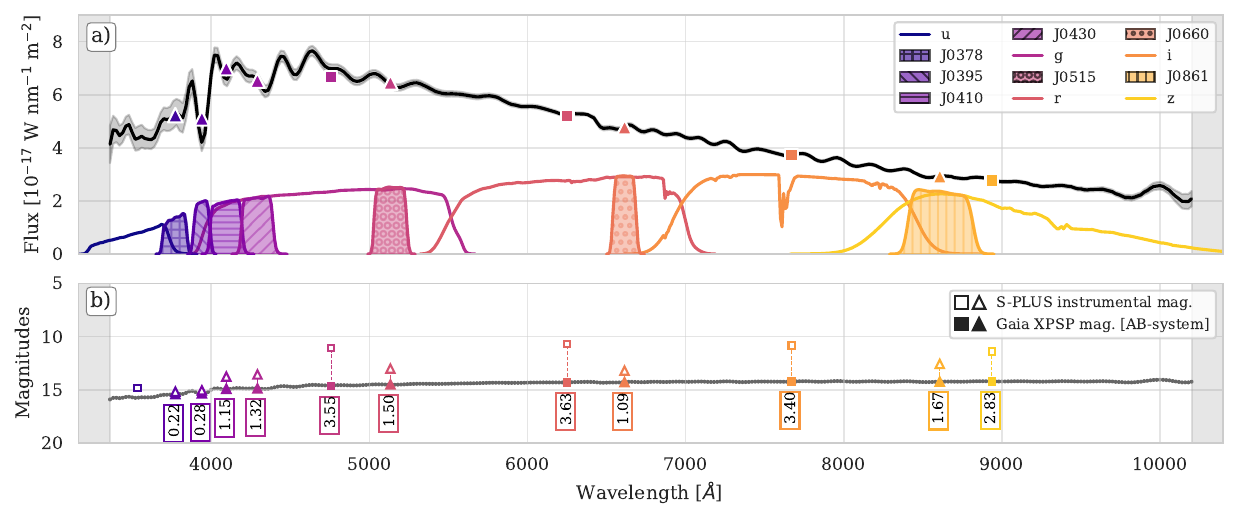}
    \caption{Comparison between \splus magnitudes and Gaia DR3 XP spectra of the source 2642100160642247424. The top panel shows a Gaia spectrum in flux (black solid line) with the \mbox{\splus} fluxes overlaid. The bottom panel shows the convoluted Gaia magnitudes compared to \splus instrumental magnitudes, where the offset in magnitudes is shown in the text boxes. The vertical gray areas represent the wavelength ranges of \splus not sampled by Gaia.}
    \label{fig:xpsp_calibration}
\end{figure*}

Synthetic magnitudes derived from these spectra serve as calibration references, replacing the external catalogs used in previous data releases. This approach has already been validated through a recalibration of \splus DR4 \citep{Xiao+2024} and \splus USS \citep{Li2025-USSRecalib}. The process, exemplified in Figure \ref{fig:xpsp_calibration} for a single star, involves generating synthetic magnitudes for hundreds to thousands of stars per field in 11 \splus filters, excluding the $u$ band, which falls outside Gaia's spectral coverage. These synthetic magnitudes are then used to calibrate the instrumental magnitudes (in the \texttt{PSTOTAL} aperture, except for the bulge, the disk, and the Magellanic Clouds, where the \texttt{PSF} aperture was used), ensuring a direct and homogeneous calibration throughout the survey footprint. In this approach, the zero-point is defined as the mode of the distribution of differences between instrumental and reference magnitudes.

For the $u$ band, which falls outside Gaia's spectral coverage, we continue to use the spectral energy distribution (SED) fitting technique employed in previous data releases. In DR5, we fit SEDs to the photometry from the 11 pre-calibrated filters to predict synthetic $u$-band magnitudes. These synthetic magnitudes then serve as the reference for determining the $u$-band zero-points.

Additionally, in previous data releases, an absolute calibration step was performed by predicting Gaia G, BP, and RP magnitudes from the calibrated \splus photometry and comparing them to observed Gaia magnitudes. This correction, which accounts for offsets of $\sim20$--$60$ mmag depending on the survey region, is no longer necessary in DR5, since the calibration is anchored to Gaia from the outset. The average offset between the predicted and the observed Gaia magnitudes is 3.8 mmag, with a standard deviation of 2.9 mmag, with no significant difference between the main survey and the fields calibrated using PSF photometry. This demonstrates the robustness of the new calibration approach.

% Data characterization
\section{Data characterization} \label{sec:data}

\subsection{Photometric Depth} \label{sec:data-depth}

The photometric depth of the images is important to understand selection effects in the catalogs and to determine the feasibility of specific analyses. Depths were determined from Petrosian magnitudes, given that this aperture is less sensitive to seeing variations and allows consistent flux estimation across different filters and objects. The analysis was based on the Main Survey dual calibration fields containing a total of \num{113496723} detections, excluding the Galactic bulge, the Galactic disk, and the Magellanic Clouds due to increased crowding in those regions, which leads to shallower depths.

For each field, depths were calculated in all 12 bands for S/N thresholds of 3, 5, 10, and 50. A kernel density estimator (KDE) from scikit-learn \citep{scikit-learn} was applied to the distribution of magnitudes at each S/N level. The depth in each case was defined as the mode of the KDE-estimated distribution, corresponding to the magnitude at which the density peaks. We adopted the KDE mode rather than the median because the magnitude distributions can be asymmetric or tailed; the mode therefore better represents the most typical limiting depth reached by the survey. The quoted upper and lower uncertainties represent asymmetric spreads computed from the 25th and 75th percentiles of the entire underlying magnitude distribution, measured relative to the KDE peak.

This analysis follows naturally from the unified structure described in the previous section, as it relies on consistent photometric measurements across all filters and apertures. Different filters exhibit varying depths due to differences in exposure time (see Table 5 of \citet{2019MNRAS.489..241M}) and throughput. Observations for DR5 were conducted under similar observing strategies, which reduces depth variation across pointings. This approach is consistent with earlier data releases.

\begin{table}[h!]
\centering
\caption{Peaks of the magnitude distributions for each band with upper and lower uncertainties, considering different $S/N$ thresholds.}
    \begin{tabular}{@{}lcccc@{}}
        \toprule
        Filter & $S/N \geqslant 50$ & $S/N \geqslant 10$ & $S/N \geqslant 5$ & $S/N \geqslant 3$ \\ \midrule
        $u$     & $17.26^{+0.25}_{-0.37}$ & $19.03^{+0.24}_{-0.44}$ & $20.00^{+0.15}_{-0.65}$ & $20.62^{+0.23}_{-0.33}$ \\
        $J0378$   & $16.78^{+0.26}_{-0.35}$ & $18.30^{+0.53}_{-0.22}$ & $19.67^{+0.10}_{-0.82}$ & $20.17^{+0.28}_{-0.31}$ \\
        $J0395$   & $16.10^{+0.48}_{-0.01}$ & $17.89^{+0.60}_{-0.24}$ & $19.17^{+0.22}_{-0.58}$ & $19.73^{+0.34}_{-0.30}$ \\
        $J0410$   & $16.11^{+0.63}_{-0.05}$ & $18.23^{+0.45}_{-0.38}$ & $19.19^{+0.35}_{-0.45}$ & $19.68^{+0.47}_{-0.19}$ \\
        $J0430$   & $16.12^{+0.69}_{-0.07}$ & $18.19^{+0.57}_{-0.23}$ & $19.03^{+0.55}_{-0.22}$ & $19.81^{+0.40}_{-0.28}$ \\
        $g$     & $18.00^{+0.26}_{-0.35}$ & $19.43^{+0.22}_{-0.24}$ & $20.54^{+0.05}_{-0.77}$ & $21.12^{+0.20}_{-0.40}$ \\
        $J0515$   & $16.79^{+0.31}_{-0.31}$ & $18.62^{+0.33}_{-0.40}$ & $19.43^{+0.21}_{-0.41}$ & $20.00^{+0.30}_{-0.28}$ \\
        $r$     & $17.99^{+0.22}_{-0.30}$ & $19.55^{+0.13}_{-0.27}$ & $20.42^{+0.15}_{-0.33}$ & $21.20^{+0.14}_{-0.30}$ \\
        $J0660$   & $17.81^{+0.14}_{-0.37}$ & $19.37^{+0.12}_{-0.34}$ & $20.29^{+0.15}_{-0.31}$ & $21.01^{+0.14}_{-0.29}$ \\
        $i$     & $17.75^{+0.12}_{-0.33}$ & $19.22^{+0.15}_{-0.22}$ & $20.01^{+0.18}_{-0.16}$ & $20.78^{+0.18}_{-0.19}$ \\
        $J0861$   & $16.77^{+0.15}_{-0.30}$ & $18.36^{+0.15}_{-0.35}$ & $19.04^{+0.22}_{-0.17}$ & $19.82^{+0.14}_{-0.23}$ \\
        $z$     & $16.97^{+0.26}_{-0.17}$ & $18.60^{+0.21}_{-0.24}$ & $19.30^{+0.16}_{-0.15}$ & $20.11^{+0.09}_{-0.31}$ \\
        \bottomrule
    \end{tabular}
\label{tab:kde_peaks_errors}
\end{table}

\subsection{Photometric Completeness} \label{sec:data-completeness}

For many science applications, knowing the completeness of samples through the parameter space can significantly affect the statistical results that can be obtained, especially when it involves counting objects. For this reason, here we provide the comparative completeness of the DR5 objects relative to those in the $r$-band, considering the detections with S/N $>5$.
 
\begin{figure*}
    \centering
    \includegraphics[width=\linewidth]{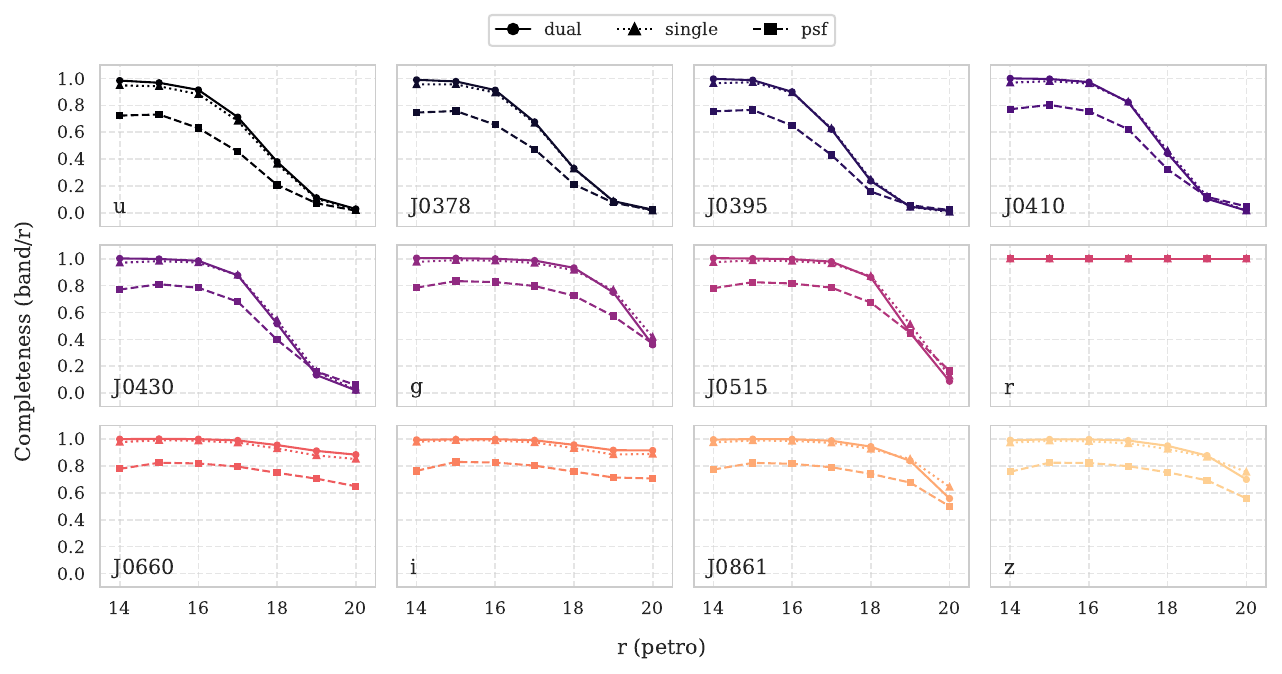}
    \caption{Completeness of source detection in different filters relative to the $r$-band, the deepest of the twelve \splus filters, at S/N$>5$. Completeness represents the fraction of sources detected in $r$ that are also detected in each one of the other filters. Each panel shows the completeness for PSF, single, and dual-mode photometry as a function of $r$ magnitude in the \texttt{PETRO} aperture.} 
    \label{fig:comp}
\end{figure*}

Figure \ref{fig:comp} shows the completeness of photometric detections as a function of the $r$-band magnitude for PSF, single, and dual photometry across all 12 \splus filters. We have chosen the $r$-band as it is the one with the highest number of detections. In each subplot of this figure, we show the completeness as the ratio between the number of sources detected in a given band and those detected in the $r$-band, for different magnitude bins.

The blue bands ($u$, $J0378$, $J0395$, $J0410$, and $J0430$) present a faster completeness drop-off, from 100\% at 15 mag to approximately 50\% at 17 mag, given the S/N threshold of 5. On the other hand, the redder bands ($J0660$, $i$, $J0861$, and $z$) have a higher completeness at fainter magnitudes, with 60-80\% at 20 mag.

% The advantage of the dual-mode photometry is clear from Figure \ref{fig:comp}, where the detection image allows both the detection of fainter objects and measurements for all filters within pre-determined apertures. For the blue filters ($u$, $J0378$, $J0395$, $J0410$, and $J0430$), the completeness drops from 100\% at 15 mag to about 50\% at 17 mag. This behavior is consistent with the one seen in DR4 photometry. 

% For the single-mode detection and PSF photometry, the completeness is also similar to that in DR4, and for the blue bands, a completeness of 50\% is reached around 18-18.5 mag.

% Catalogs
\section{Catalogs} \label{sec:cat}

\subsection{Data Structure} \label{sec:cat-structure}

DR5 adopts a more consolidated data structure compared to previous releases. Unlike DR4, where each filter had a dedicated table, DR5 organizes all photometric measurements into unified tables, simplifying data access. Moreover, the set of columns remains the same as described in \cite{2024A&A...689A.249H}.

The data are distributed across two main schemas:
\begin{itemize}
    \item \texttt{dr5}: Contains the core photometric measurements, including single-mode, dual-mode, and PSF photometry, each provided in a dedicated table: \texttt{dr5\_single}, \texttt{dr5\_dual}, and \texttt{dr5\_psf}, respectively. 

    \item \texttt{dr5\_vacs}: Hosts value-added catalogs (VACs), which provide derived parameters and enhanced data products, as detailed in the next section.
\end{itemize}

Each table in the \texttt{dr5} schema integrates measurements for all \splus filters --- $u$, $J0378$, $J0395$, $J0410$, $J0430$, $g$, $J0515$, $r$, $J0660$, $i$, $J0861$, and $z$ --- within a single structure. In addition to fluxes and magnitudes, each table also includes morphological parameters (e.g., the full width at half-maximum, ellipticity, isophotal area) for a given source and filter.

A complete list of columns and their descriptions for each table is provided in the DR5 documentation\footnote{\url{https://splus.cloud/documentation/dr5}}.

\subsection{Value-Added Catalogs (VACs)} \label{sec:cat-vacs}

\subsubsection{Star-quasar-galaxy Classification (SQG)} \label{sec:cat-sqg}

Compared to DR4, several improvements have been implemented in the photometric classification process for this \splus DR5 VAC. The star-quasar-galaxy classification model that uses \splus magnitudes, Wide-field Infrared Survey Explorer \citep[WISE;][]{Wright2010-WISE} magnitudes, and morphological parameters has been retrained using the new DR5 data. This model, referred to as the Base Model, maintains continuity with the version employed in DR2, DR3, and DR4, and further details are available in the documentation and in Nakazono et al. (in preparation).

In previous releases, two separate models were trained for sources with and without WISE counterparts. In DR5, however, we adopt a single unified model. Objects lacking a WISE detection are assigned an out-of-range placeholder value of 99 during training, allowing the model to handle both cases simultaneously. As a result, the column \texttt{model\_flag} is no longer provided in DR5. This modification was motivated by the improved overall performance observed when using a unified approach \citep{sibgrapi_estendido}. However, the expected classification accuracy remains higher for sources with WISE detections, as discussed in \citet{Nakazono2021}.

Another major update is the replacement of the AllWISE \citep{Wright2019-AllWISE} catalog with the unWISE \citep{Dustin2014-unWISE} catalog, which offers improved photometry and superior source deblending. This enhancement is expected to lead to more reliable classifications, particularly in crowded regions and for faint sources.

In addition to these improvements, we developed a new classification model that incorporates Gaia EDR3 \citep{Gaiaedr32021} parameters. This Gaia Model includes information such as parallax, proper motion, and $G$-band magnitude, providing enhanced discrimination between stars, quasars, and galaxies. The outputs of this model are reported in separate columns in the VAC, distinguished by the suffix \texttt{\_GAIA}: \texttt{CLASS\_GAIA}, \texttt{PROB\_STAR\_GAIA}, \texttt{PROB\_QSO\_GAIA}, and \texttt{PROB\_GAL\_GAIA}. External parameters used in the classification, including the unWISE and Gaia data (\texttt{W1\_MAG}, \texttt{W2\_MAG}, \texttt{Gmag}, \texttt{Plx}, \texttt{E(BP/RP)}, and \texttt{PM}), are also provided in the VAC for completeness and reproducibility.

Considering a magnitude limit of $r < 21$ and using the \textit{Gaia Model}, the contamination rate is below 2\% for quasars, and less than 1\% for stars and galaxies. The corresponding miss rates (i.e., fraction of misclassified sources) are approximately 2\% for quasars, less than 0.5\% for stars, and around 0.6\% for galaxies.

\subsubsection{Photometric Redshifts (photo-zs)} \label{sec:cat-pz}

The photometric redshifts for the DR5 are obtained using a Bayesian Mixture Density Network \citep[BMDN;][]{MDN1994} machine learning model. This model is capable of estimating precise and accurate single-point estimates (SPEs) while also providing well-calibrated probability density functions (PDFs) \citep{2022A&C....3800510L}.

The model is trained using photometry from \splus data crossmatched with the Galaxy Evolution Explorer \citep[GALEX;][]{Martin2005-GALEX}, Vista Hemisphere Survey \citep[VHS;][]{McMahon2021-VHS}, and unWISE survey data. To obtain the spectroscopic redshifts (spec-zs) needed, we crossmatch \splus DR5 with a compilation of spec-zs in the Southern Hemisphere \citep{erik_zenodo}. The training dataset contains \num{867636} galaxies in the magnitude range between 14 and 21 in the \texttt{r\_auto} band (corrected for extinction using the Corrected SFD dust map \citep{Chiang2023-CSFD} and the \citet{CCM89} extinction law) and between 0.002 and 0.8 in redshift.

% The model is evaluated using four metrics for single-point estimates: the normalized median absolute deviation \citep[$\sigma_\text{NMAD}$;][]{Brammer2008}, bias, normalized bias, and outlier fraction \citep{Ilbert2006}. It is also evaluated using six other metrics for the probability density functions: Probability Integral Transform and Continuous Ranked Probability Score, as defined in \citet[PIT and CRPS;][]{Polsterer2016}, the Highest Probability Density Credible Interval \citep[HPDCI;][]{HPDCI,Wittman2016_HPDCI2}, the Odds \citep{Benitez2000}, the 1-sigma confidence interval ($\sigma_{68}$), and the PDF maximum height. Further details can be found in \cite{2022A&C....3800510L}. 

The model is evaluated using four metrics for single-point estimates: the normalized median absolute deviation \citep[$\sigma_\text{NMAD}$;][]{Brammer2008}, bias, normalized bias, and outlier fraction \citep{Ilbert2006}. It is also evaluated using six other metrics for the probability density functions: 1) Probability Integral Transform and 2) Continuous Ranked Probability Score, as defined in \citet[PIT and CRPS;][]{Polsterer2016}; 3) the Highest Probability Density Credible Interval \citep[HPDCI;][]{HPDCI,Wittman2016_HPDCI2}; 4) the Odds \citep{Benitez2000}; 5) the 1-sigma confidence interval ($\sigma_{68}$); 6) and the PDF maximum height. Further details can be found in \cite{2022A&C....3800510L}.

The metrics obtained for the entire sample between $r$-band magnitudes of 14 and 21, and in the redshift range of 0.002 to 0.8, are a scatter of 0.0248, a bias of $-5.03\times 10^{-4}$, and an outlier fraction of 1.44\%.

The metrics obtained for five magnitude bins of equal population size are shown in Table \ref{tab:pz_perf}. The bin edges were defined so that each bin contains 20000 objects:
\begin{table}[h!]
\caption{Photometric redshift performance metrics (scatter, bias, and outlier fraction) in bins of the r-band magnitude, corrected for extinction.}
\hspace{-0.7cm}
    \begin{tabular}{@{}lccc@{}}
        \toprule
        Magnitude bin  & $\sigma_\text{NMAD}$ & $\mu$ $(\times 10^{-4})$ & $\eta$ (\%) \\ \midrule
        14.00 to 18.05 & 0.0106               & -1.03                    & 0.18        \\
        18.05 to 18.98 & 0.0208               & -4.20                    & 0.57        \\
        18.98 to 19.55 & 0.0282               & -5.81                    & 0.99        \\
        19.55 to 20.07 & 0.0349               & -1.18                    & 1.74        \\
        20.07 to 21.00 & 0.0424               & -35.6                    & 3.71        \\ \bottomrule
    \end{tabular}
    \label{tab:pz_perf}
\end{table}

%The photo-zs whose metrics are described in this paper correspond to the table \texttt{dr5\_photozs\_v3}, which provide good photo-z performance over the full redshift range ($0 \leq z \leq 0.8$). In addition, a version optimized for low-redshift galaxies is also available and described in Mendes de Oliveira et al. (submitted) and Placco et al. (in prep.).

\subsubsection{Bright Star Masks (BSM)} \label{sec:cat-bsm}
To mitigate contamination from bright stars in the \splus DR5 fields, we generate masks using the Mangle\footnote{\url{https://space.mit.edu/~molly/mangle/}} software \citep{Swanson2008-Mangle}. For the masks, the region to be assigned a zero weight is approximated to be circular around the center of the saturated star. These masks are designed to identify and exclude areas affected by saturated stars and their associated artifacts, such as spikes and halos.

Two complementary methods were developed to define the circular region. The first algorithm identifies groups of saturated pixels in \splus images by first normalizing the pixel values to the [0,1] range and applying a fixed threshold of 0.55 (defined after visual inspection) to isolate saturated pixels. These pixels are then clustered using the Density-Based Spatial Clustering of Applications with Noise \citep[DBSCAN;][]{Ester1996-DBSCAN} algorithm based on spatial proximity. For each cluster of pixels, the star's position is estimated via an intensity-weighted average of the pixel coordinates, and a circular mask radius is defined as 60\% of the maximum pairwise distance within the cluster, plus a fixed minimum of 30 pixels. In general, the masked pixels represent a few percent of the total area. 

However, this method is not fully effective in fields with extremely bright stars. Therefore, we apply a second method that uses the HST Guide Star Catalog \citep[GSC, ][]{2001AJ....121.1752M} to identify stars with \mbox{$P_{\text{mag}} < 14$} mag within a one-degree radius from each field center. This second method is also triggered when a field does not present masked pixels beyond the 0.55 threshold. A circular exclusion radius is calculated for each star using the empirically calibrated formula (Eq. \ref{eq:radius}):
\begin{equation}
    \label{eq:radius}
    r = 19223 e^{-(P_{\text{mag}} + 11.1)^2 / 76.88}.
\end{equation}

The radius assigned to a given star was set to extend as far from its center as necessary for the brightness to reach the background level. A minimum exclusion radius of 30 pixels (16.5 arcsec) is set to avoid the circle reaching $r=0$ for stars at the faint end ($P_{\text{mag}} \rightarrow 14$). During our tests, we found that approximately 1\% of the \splus saturated stars are not present in the GSC catalog; thus, we prioritize the first algorithm. 

We provide the masked regions in \texttt{Mangle} format\footnote{Available for download here [will be added after the report process.]}. With these masks, we also create a VAC with a boolean value to identify any object that is within the radius calculated. A value of \texttt{is\_clean = True} indicates that the source is in a valid region, \texttt{is\_clean = False} otherwise.

We stress that objects with \texttt{is\_clean = True} are not guaranteed to be completely free from contamination. The object can still be near a stellar spike while outside the ``saturation radius'' and, therefore, contaminated to some extent. Such features are intrinsically difficult to model, and there is ongoing work within the technical team to develop a tool that can tackle such issues. However, objects with \texttt{is\_clean = False} are statistically more likely to be contaminated and, therefore, must be carefully considered or excluded from analysis. Using the \texttt{SExtractor} quality flags\footnote{\url{https://sextractor.readthedocs.io/en/latest/Flagging.html}} to get an estimate of saturated and non-saturated sources, we found a false positive rate of 0.12\% (using $flag = 0$) and a median false negative rate of 1.63\% (using $flag \geqslant 4$). 

\subsubsection{Extinction Coefficients (EXT)} \label{sec:cat-ext}

This value-added catalog is a new addition in DR5 and is meant to simplify the process of obtaining interstellar medium (ISM) extinction corrected magnitudes for the \splus sources.
The data provided in the main DR5 catalogs are not corrected for ISM extinction. For the application of the corrections, we provide the coefficients ($A_\lambda$) for each \splus filter based on three sources: the $E(B-V)$ maps from \citet{Schlegel1998}, labeled as \texttt{SFD}; the 86\%-scaled \texttt{SFD} correction proposed by \citet{Schlafly2011}, labeled as \texttt{SFLY}; and Gaia DR3, where the coefficients are based on the measured Gaia $G$-band extinction ($A_G$) and labeled as \texttt{GAIA3}.

To estimate $A_\lambda$, we adopted $R_V = 3.1$ and the $A_\lambda/A_V$ coefficients derived for the \splus filters assuming a \citet{Fitzpatrick1999} dust law. The extinction coefficients are provided in \citet[][Table 3]{Perottoni2024} and described in detail in \citet[][Appendix A]{2024A&A...689A.249H}.

Representing the AB magnitude of a given filter in a given aperture as $m$, the ISM extinction corrected magnitude ($m_0$) can be obtained using Equation \eqref{eq:extinction}:
\begin{equation}
m_0 = m - A_\lambda
\label{eq:extinction}
\end{equation}

\subsubsection{Overlapping Region Flags (ORF)} \label{sec:cat-orf}

The \splus observation strategy generates a certain overlap between adjacent tiles. Although this is not an issue when analyzing individual fields, it can induce systematic biases when combining adjacent fields or studying large contiguous regions, particularly in analyses sensitive to number density or spatial distributions. These overlaps arise because objects located in overlapping regions are observed multiple times and may appear as duplicate detections. For this reason, we introduce a new VAC in DR5 that contains a flag that indicates whether the object is in the overlap region or not.

\begin{figure}
    \centering
    \includegraphics[width=\linewidth]{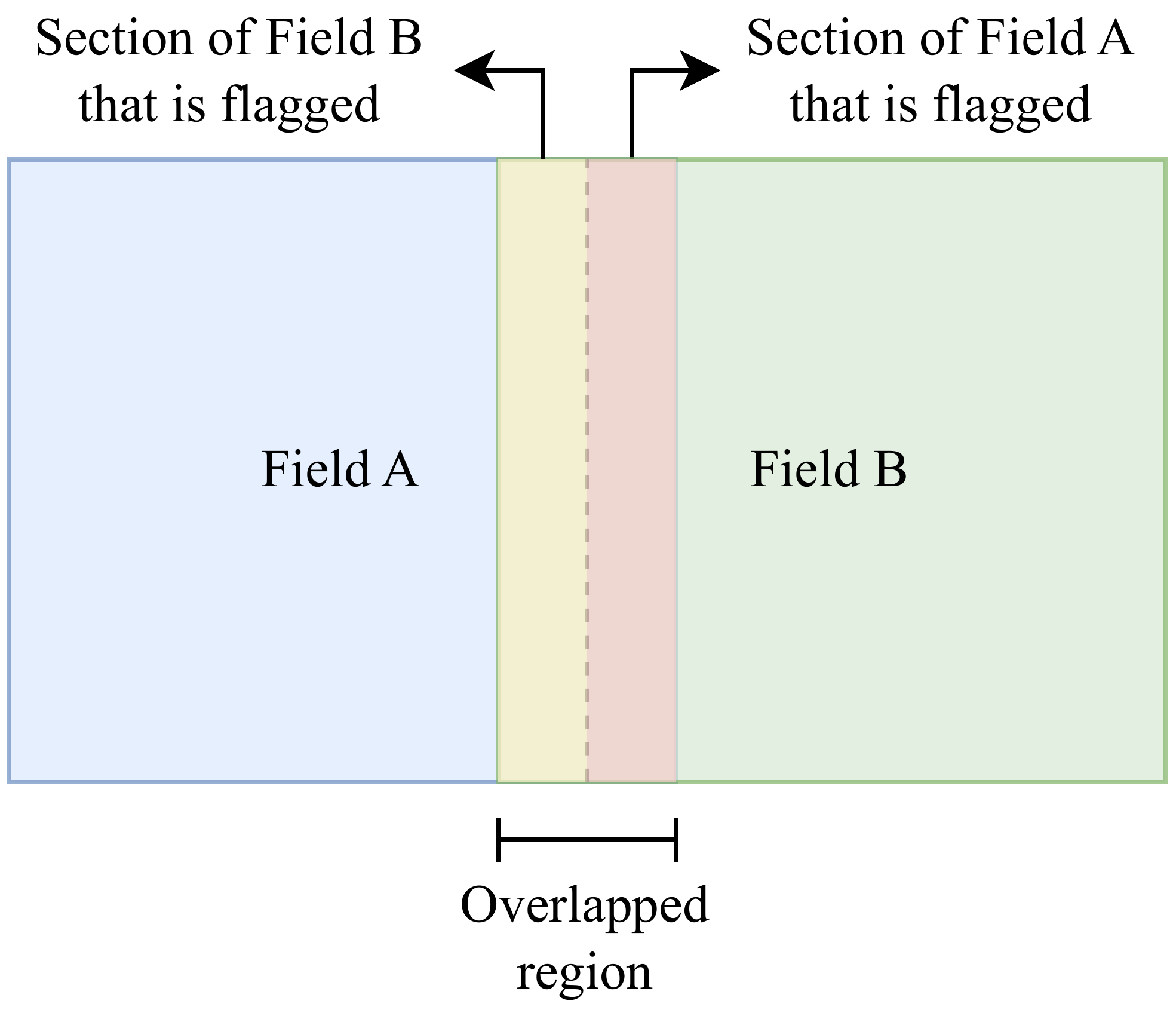}
    \caption{Illustration of the overlap flagging process.}
    \label{fig:overlap_explain}
\end{figure}

\begin{figure*}
    \centering
    \includegraphics[width=\linewidth]{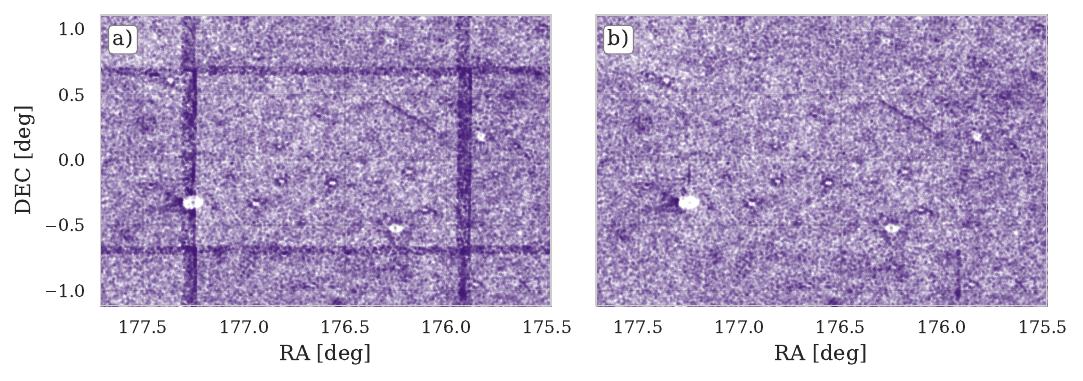}
    \caption{Comparison of the field ``SPLUS-n01s20'' and surrounding fields before (a) and after (b) applying the overlap flag condition.}
    \label{fig:border_flag}
\end{figure*}

This flag enables users to identify and properly handle duplicated sources when combining adjacent fields. The flag is assigned using a simple and robust algorithm that compares each field with its adjacent fields in pairs. Taking Figure \ref{fig:overlap_explain} as an example, we compare Field A to Field B. If there is an overlap, we flag the farther half of the overlap relative to Field A, which is represented by the red region in the figure. Afterwards, we reverse the comparison and compare Field B to Field A using the same process, and flag the yellow region. This bidirectional approach ensures that overlapping regions are consistently identified in both fields, preventing biases in regions of higher source density, as shown in Figure \ref{fig:border_flag}.

To remove objects inside the overlap, one should select only those with \texttt{in\_overlap\_region} = 0. Note that we do not apply any quality constraints when selecting the objects to be flagged. Thus, using \texttt{in\_overlap\_region} = 0 does not guarantee that a particular selected source has the best $S/N$ or meets any other quality criteria. Therefore, we strongly recommend that users evaluate the impact of this flag on their specific science case before blindly removing sources. % so we recommend prudence when using this flag for some types of science

\subsubsection{\splus Unique ID} \label{sec:cat-suid}

To ensure consistent identification of sources across different catalogs and photometric modes in DR5, we provide a VAC containing a universal identifier, referred to as the \texttt{unique\_id}, for each \splus object. These identifiers are generated from their sky coordinates (RA, Dec), encoded in the ICRS frame following a fixed formatting scheme.

The \texttt{unique\_id} is created from the object's sky position and then propagated by positional matching. For each detection, RA and Dec are converted to fixed-format sexagesimal strings and concatenated with the prefix \splus, forming an identifier such as \texttt{SPLUS100029.63-303915.6}. Inside a field, detections within 3 arcseconds are matched and forced to share the same \texttt{unique\_id}. Between neighboring fields, detections are cross-matched using a radius that depends on \texttt{CLASS\_STAR}, and if a match exists with a non-zero \texttt{unique\_id}, that identifier is copied. This ensures that the same object observed in different fields receives the same \texttt{unique\_id}. Among duplicates that share the same \texttt{unique\_id}, only the detection with the highest $S/N$ is flagged for selection. This flag should be interpreted as a recommended default for constructing a single-entry catalog, not as a mandatory filter. Since duplicate detections are preserved in the database, users may instead build alternative selections, including weighted combinations or other quality-based choices.

This unique identifier enables consistent merging of catalogs, deduplication, and unambiguous cross-referencing of sources in large-scale analyses.

\subsection{Data Access} \label{sec:cat-access}

All \splus data, including previous data releases, are hosted on the collaboration server\footnote{\url{https://splus.cloud}} \citep{oliveira_schwarz_2022_10980447}. In the ``Tools'' section of the website, users can access various tools for different tasks as previously described in \citet{2024A&A...689A.249H}.

The website integrates the Table Access Protocol (TAP), established by the International Virtual Observatory Alliance (IVOA), as described in \citet{2017ivoa.spec.0509L}. TAP allows users to execute \textit{ad hoc} queries using the SQL-like Astronomical Data Query Language (ADQL) on the stored data tables and apply geometrical constraints.

Furthermore, the Python \texttt{splusdata}\footnote{\url{https://github.com/Schwarzam/splusdata}} package provides an interface for accessing \splus data programmatically. This package enhances usability, enabling integration into different workflows.
A detailed guide for accessing the data is available on the DR5 documentation page\footnote{\url{https://splus.cloud/documentation/dr5}}.

% Science application
\section{Representative Science Application} \label{sec:science}

As previously mentioned in Section \ref{sec:intro}, the new \splus DR5 data provide homogeneous imaging of the Antlia and Hydra clusters, opening a valuable window on galaxy evolution in nearby dense environments. Both systems lie at $z\sim0.01$ \citep{Korteweg1995,Castelli2008} and are part of the Hydra Supercluster. This makes the Hydra–Antlia region especially attractive: it offers the opportunity to characterize the assembly and environmental processing of galaxies in a low-richness supercluster, providing a complementary perspective to the well-studied, richer systems. On even larger scales, the Hydra Supercluster and its neighbor, the Centaurus Supercluster, have been associated with the Laniakea region, and appear to be connected to the Virgo Supercluster through the surrounding network of filaments \citep{2014Tully}. The proximity of Antlia and Hydra allows us to resolve faint galaxy populations, up to B$_\text{T}=22.6$ mag, or central surface brightness $\mu_{0,g} \geqslant$ 24 mag.arcsec$^{-2}$ \citep{2012MNRAS.419.2472S, Iodice2023-FaintHydra, Buttitta2025-FaintHydra}, and accurately study their properties using the \splus 12-band photometric system. In the following sections, we use a robust spectroscopic subsample to calibrate our \splus photometric redshifts (photo-zs) and identify cluster members with the ``shifting-gapper'' technique \citep{Fadda1996,Lopes2009}. Using the identified members, we map projected density contours across the Hydra–-Antlia area and utilize the narrowband $J0660$ filter to trace star-forming galaxies as a function of cluster-centric distance. The original coordinates and redshifts of our systems come from the list of \citet{Yuan2020-Sample}\footnote{\url{http://zmtt.bao.ac.cn/galaxy_clusters/dyXimages/combined.html}}. To ensure a robust dynamical characterization and maximize the symmetry of the velocity distribution, we refined the central coordinates and redshifts by re-estimating them from member galaxies within $0.30~h^{-1}$ Mpc. For all results shown, we adopt $H_0 = 70$~km~s$^{-1}$~Mpc$^{-1}$, $\Omega_m = 0.3$, and $\Omega_\Lambda = 0.7$.

\subsection{Photometric Cluster Membership and\\Large Scale Structure} \label{sec:membership_lss}

To fully exploit the \splus photometric data for cluster membership, we must first establish a robust spectroscopic baseline to calibrate our photometric redshifts. To identify the spectroscopic members, we apply the ``shifting gapper'' technique. This approach combines galaxy line-of-sight velocities and cluster-centric distances to efficiently suppress interlopers. First, we select a subsample within 0.5$h^{-1}$ Mpc of the cluster center. Within this region, we obtain estimates of the minimum and maximum velocity of galaxy members, which we call $v_\text{low}$ and $v_\text{high}$, respectively. Those are used to compute the two limits of the velocity offsets relative to the cluster redshift. We then call the maximum absolute value of these two as $v_\text{diff}$:
\begin{align*}
    v_\text{diff} = \max(|v_\text{low}-v_\text{cluster}|, |v_\text{high}+v_\text{cluster}|).
\end{align*}

In order to have a symmetric velocity window with respect to the cluster velocity, we update the values obtained above, now having $v_\text{low} = v_\text{cluster} - v_\text{diff}$ and $v_\text{high} = v_\text{cluster} + v_\text{diff}$. We also compute the parameter $facgap = min(300, |v_\text{low} - v_\text{high}|/10)$. These values are used to define the velocity-gap threshold: 
\begin{align*}
    \Delta v = facgap(1+\exp(-(N-6)/33)),
\end{align*}
where $N$ is the number of members within each radial bin (see below).

Following this procedure, we bin galaxies with respect to their cluster-centric distance in bins of 0.42$h^{-1}$ Mpc, adaptively adjusted to contain at least 15 galaxies per bin. We then apply a velocity-gap test, where the velocity of the galaxies are sorted, and the difference between consecutive values (i.e., the gap) is calculated. Galaxies with velocity gaps below the $\Delta v$ threshold are marked as candidate members, while the others are considered interlopers. This procedure is iteratively repeated outward until the radial offset between two consecutive galaxies is higher than 0.75 sep$_\mathrm{biw}$ (where  sep$_\mathrm{biw}$ is the biweight estimate of the typical separation of the central galaxies, within 0.5$h^{-1}$ Mpc; \citealt{Beers1990Biweight}). Within each radial bin we update the values of $v_\text{low}$, $v_\text{high}$ and $facgap$. We continue this process until no more galaxies are removed, obtaining a list of member candidates and interlopers for the clusters, which are used in the subsequent analysis.

The maximum radius we sample is 10$h^{-1}$ Mpc (as we want to cover the infall region of the clusters). However, to derive an initial estimate of the velocity dispersion ($\sigma_\text{cl}$), we restrict the analysis to members within $2.50~h^{-1}$ Mpc, which is larger than the physical radius ($R_{200}$) of massive clusters. An initial estimate of the virial mass, $M_{200}$, and radius, $R_{200}$, can be obtained as described in \citet{Lopes2024}, and the final values consider only galaxies within $R_{200}$.

\begin{figure*}[ht]
    \centering
    \includegraphics[width=\linewidth]{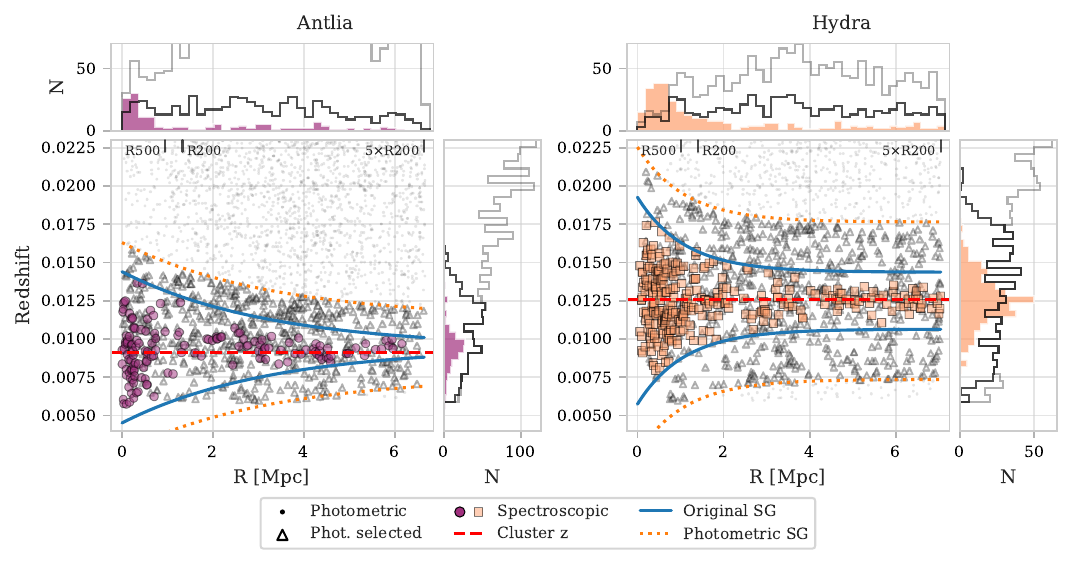}
    \caption{Redshift distribution of member galaxies (both spectroscopic and photometric) as a function of distance to the cluster center, within $5\times R_{200}$. The left panels shows 745 members of Antlia ($z = 0.00976$), and the right panels shows 961 members of Hydra ($z = 0.01252$). Black triangles represent the photo-z objects selected using our approach, shifted by 0.0022 for Antlia and 0.0027 for Hydra to align with the spec-zs (see text). The vertical lines show the $R_{500}$ (the radius at 500 times the critical density of the Universe), $R_{200}$, and $5\times R_{200}$ distances. The small gray dots represent photometric redshift objects not selected.}
    \label{fig:hist_ant_hyd}
\end{figure*}

In Figure \ref{fig:hist_ant_hyd}, we present the distribution of spectroscopic redshifts of member galaxies as a function of their distance from the cluster center, inside $5\times R_{200}$ ($R_{200} \sim 1.4$ Mpc for each cluster). For Antlia (in purple, left panel), we identify 151 members, including those in \cite{2015A&A...584A.125C} and \cite{2015MNRAS.451..791C}, also included in the Southern Hemisphere Spectroscopic Redshift Compilation \citep{erik_zenodo}. For Hydra (in orange, right panel), we identify 311 members. Using our identified members, we can infer the physical properties of both clusters. For Antlia, we obtain an average $z$ of 0.00976, a velocity dispersion ($\sigma_\text{cl}$) of 646.11~km~s$^{-1}$, a $M_{200}$ of 2.70$\times10^{14}$ $M_\odot$ and $R_{200}$ of 1.33 Mpc. For Hydra we get 0.0125, 686.74~km~s$^{-1}$, 3.22$\times10^{14}$ M$_\odot$, and 1.41 Mpc, respectively.

For the photo-zs, we apply a bias correction procedure (Mendes de Oliveira et al., in prep) to align the spec-z and photo-z distributions. This procedure consists of Gaussian fits to the histogram of both the photo-z and spec-z distributions in the region of the cluster and calculating the difference (shift) between the mean of these Gaussians. We apply this measured shift as a global correction to the photometric redshifts, resulting in offsets of 0.0022 for Antlia and 0.0027 for Hydra.

\begin{figure*}[ht]
    \centering
    \includegraphics[width=\linewidth]{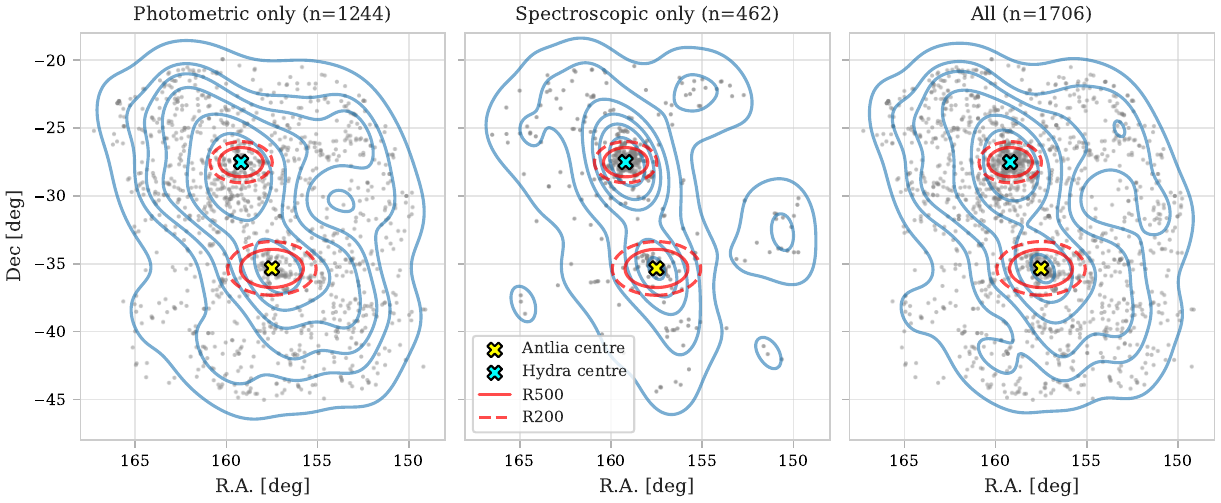}
    \caption{Density contour plot for the galaxies in the region of Antlia and Hydra. The left panel shows the new photometrically determined members exclusively with photo-zs, the central panel shows members determined with spec-zs, and the right panel shows both simultaneously. The red solid and dashed circles represent the $R_{500}$ and $R_{200}$ of the clusters, respectively. In both panels, we observe a bridge between the centers of Antlia and Hydra, which forms the Hydra-Antlia wall.}
    \label{fig:contour_hydra_antlia}
\end{figure*}

By combining the calibrated \splus photo-zs with the spectroscopic sample, we can map the extended structure of the clusters far beyond the spectroscopic limits. Figure \ref{fig:contour_hydra_antlia} shows the density contours for the galaxies in the Hydra and Antlia clusters, selected by photo-z on the left panel, and by spec-z on the middle panel. The cut in photo-z was defined as $z = 0.0111^{+0.015}_{-0.005}$ (i.e., the average redshift of both clusters) and any object with photo-z $odds$ $\geqslant 0.6$ and within a ``broadened'' shifting gapper envelope has been considered a member of the clusters. The $odds$ is a metric related to how concentrated a PDF is around its peak. A value of 1 means that the entire PDF falls inside the interval of photo-z $\pm 0.02$, as described in \citet{Benitez2000}. The ``broadening'' is done using the photo-z of the spectroscopic members, and the upper and lower curves of the envelope are shifted to contain at least 90\% of these galaxies. The photometric objects selected with this approach are shown in Figure \ref{fig:hist_ant_hyd} as black triangles, while rejected photometric members are shown as gray dots. It is noteworthy that the \splus photometric data allow us to identify 1244 new cluster members, as seen in the left panel of Figure \ref{fig:contour_hydra_antlia}. This increases the total known membership by roughly $269\%$ (from 462 to 1706 objects).

\subsection{\splus multiband color analysis} \label{sec:color_analysis}

The color of galaxies, as calculated from the difference of magnitudes in two photometric bands (using the \texttt{AUTO} aperture in this work) provide insight into their formation, evolution, and environment. 

\begin{figure*}[ht]
    \centering
    \includegraphics[width=\linewidth]{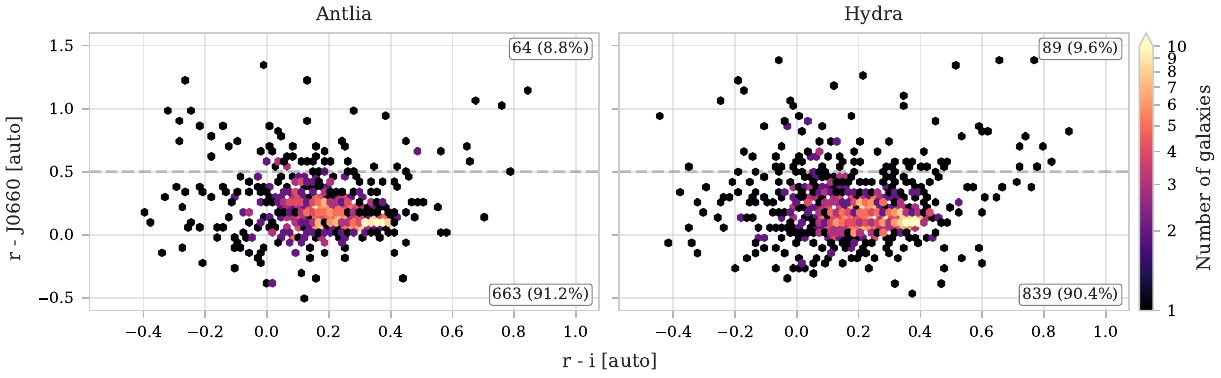}
    \caption{Color-color diagram of ($r-i$) versus ($r-J0660$) for all galaxies within $5\times R_{200}$ from the center of Antlia (left) and Hydra (right). The points are colored according to the density of objects. The inset numbers quantify both the total number of galaxies and the fraction of objects above the adopted  $(r-J0660)=0.5$ threshold, highlighting the relative contribution of star-forming systems in each cluster.}
    \label{fig:color-color}
\end{figure*}

Beyond standard broadband colors, the \splus 12-band system provides a unique window into the ionized gas content of these galaxies. Specifically, the narrowband $J0660$ filter covers the H${\alpha}$ and [\ionn{N}{ii}] emission lines at the redshift of these clusters. In Figure \ref{fig:color-color}, we show a diagram of (\mbox{$r-J0660$}) versus (\mbox{$r-i$}) for all galaxies (photometric and spectroscopic) that are located within $5\times R_{200}$ of the center of each cluster. This allows us to identify galaxies with significant H$\alpha$ excess \citep{GutierrezSoto2025,Lopes2025}, which is a direct tracer of recent star formation activity \citep{Hopkins2003-SFR,Kramarenko2026-SFR}. It can be seen that both clusters contain a similar fraction of this kind of galaxy, amounting to approximately 9\% of the total number of objects in each. A catalog of 80 visually inspected H$\alpha$-emitting galaxies within $5\times R_{200}$ of the center of the Hydra cluster using \splus DR5 data is published in Cardoso et al. (in prep).

\begin{figure*}[ht]
    \centering
    \includegraphics[width=\linewidth]{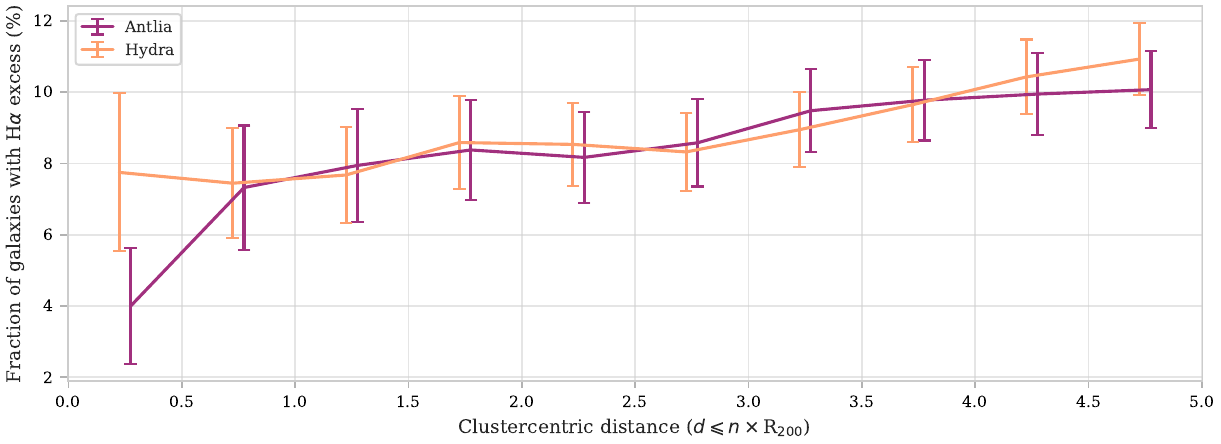}
    \caption{Comparison of the fraction of H$\alpha$-excess galaxies between Antlia and Hydra up to $5\times R_{200}$. Both have similar increasing trends for higher distances, whereas Hydra has a larger fraction of galaxies with excess H$\alpha$ at its center.}
    \label{fig:color-color_distance}
\end{figure*}

In Figure \ref{fig:color-color_distance}, we analyze the H$\alpha$-excess galaxies identified in Figure \ref{fig:color-color} (using the \mbox{$r-J0660 \geqslant 0.5$} threshold). In this plot, we use bootstrapping to estimate uncertainties and analyze the fraction of this kind of galaxy as a function of its respective clustercentric distance. We find that both clusters possess similar fractions of H$\alpha$-excess galaxies, exhibiting an increasing trend toward higher clustercentric distances. Furthermore, Hydra contains a higher fraction of these galaxies at its core, with a fraction of $7.74 \pm 2.22\%$ compared to $4.00 \pm 1.62\%$ for Antlia at distances below $0.5\times R_\text{200}$. We note that this color–color approach is effective at detecting the most intense emitters, such as galaxies with high star formation rates or active nuclei. A more complete census of H$\alpha$ emitters requires a continuum subtraction, either using a Three-Filter Method \citep{Lopes2025} or via a SED fitting technique \citep{ThainaBatista2023}.

% Conclusions
\section{Conclusions}
\label{sec:conclusion}

In this paper, we presented the fifth data release (DR5) of \splus, increasing our coverage of the Southern Hemisphere sky from $\sim$3000 to $\sim$4592 square degrees, including $\sim$110 square degrees of coverage of the Galactic bulge and disk.

We have shown the main technical and scientific advancements for this release, including the introduction of the new reduction pipeline, the Multiband Astronomical Reduction (MAR), which was used to reduce over 2200 fields with significantly improved astrometric precision, allowing us to process, for the first time, fields in the Galactic bulge and disk (Section \ref{sec:obs-reduction}). It delivers point-spread-function (PSF) photometry for 371 fields located in the bulge and disk, while aperture photometry is provided for the remaining fields, in both single- and dual-mode catalogs. These modes correspond to independent detections in individual bands and to combined detections using the $griz$ bands, respectively (Section \ref{sec:obs-photometry}).

For this new data release, we also implemented a new photometric calibration strategy based on Gaia DR3 XP spectra, offering uniform calibration throughout the footprint and naturally accounting for interstellar extinction. This improves both the precision and reliability of \splus photometry (Section \ref{sec:obs-phot_calibration}). In conjunction with the MAR pipeline, this allows us to reach a magnitude depth of 21.5 in the $r$ band for objects with S/N $\geq 3$, and 19.55 for those with S/N $\geq 10$ (Section \ref{sec:data-depth}).

We also provide descriptions for the VACs in this release in Section \ref{sec:cat-vacs}. Those include the star-galaxy-quasar (SGQ) separation, the photometric redshifts (photo-zs), and four new catalogs: bright star masks (BSMs), overlapping region flags (ORFs), extinction coefficients (EXTs), and \splus Unique IDs. Changes in both SGQ and photo-zs methodologies and the increase of available data translated into improved results when compared to previous data releases. Additionally, the BSMs allow filtering objects that can have affected photometry due to proximity to bright stars, while ORFs permit removing overdensities near the border of fields, providing improved capability for density studies. Finally, the EXTs facilitate the usage of \splus magnitudes by providing the coefficients for extinction correction, and the unique IDs allow crossmatching the new data with previous and future releases.

To showcase the capability of \splus observations, we presented a science case based on the detection and characterization of members of the Hydra and Antlia clusters in Section \ref{sec:science}. This is noteworthy since the filter system of the survey was designed primarily for the study of stars, but can also yield improved results for extragalactic cases, given the high-precision derived photometric redshifts. We expanded the number of members of the clusters from 462 to 1706, incorporating both photometric and spectroscopic redshifts, and found evidence that supports the existence of a ``bridge'' connecting both structures. Additionally, we found that the Hydra cluster has a higher proportion of H$\alpha$-excess galaxies in its center when compared to Antlia.

All the data presented in this paper and from the previous data releases are available on the collaboration server (\url{https://splus.cloud}) and via the \texttt{splusdata} Python package.

% Acknowledgements
\begin{acknowledgments}
\label{sec:acknowledgements}

We are grateful to the referee that provided useful comments and insights during the revision process of this paper. 

The author acknowledges the financial support given by Coordenação de Aperfeiçoamento de Pessoal de Nível Superior (CAPES, grant 88887.470064/2019-00), Conselho Nacional de Desenvolvimento Científico e Tecnológico (CNPq, grant 169181/20170), and Fundação de Amparo à Pesquisa do Estado de São Paulo (FAPESP, grant 2024/15229-8) during the development of this research. We also thank Vasiliki Fragkou and Alvaro Alvarez-Candal for their careful read and comments on this paper.

The S-PLUS project, including the T80-South robotic telescope and the S-PLUS scientific survey, was founded as a partnership between the Fundação de Amparo à Pesquisa do Estado de São Paulo (FAPESP), the Observatório Nacional (ON), the Federal University of Sergipe (UFS), and the Federal University of Santa Catarina (UFSC), with important financial and practical contributions from other collaborating institutes in Brazil, Chile (Universidad de La Serena), and Spain (Centro de Estudios de Física del Cosmos de Aragón, CEFCA). We further acknowledge financial support from the FAPESP grant 2019/263492-3, the Brazilian National Research Council (CNPq), the Coordination for the Improvement of Higher Education Personnel (CAPES), the Carlos Chagas Filho Rio de Janeiro State Research Foundation (FAPERJ), and the Brazilian Innovation Agency (FINEP). The S-PLUS collaboration members are grateful for the contributions from CTIO staff in helping in the construction, commissioning, and maintenance of the T80-South telescope and camera. We are also indebted to Rene Laporte, INPE, and Keith Taylor for their essential contributions to the project. From CEFCA, we particularly would like to thank Antonio Marín-Franch for his invaluable contributions in the early phases of the project, David Cristóbal-Hornillos and his team for their help with the installation of the data reduction package JYPE version 0.9.9, César Íñiguez for providing 2D measurements of the filter transmissions, and all other staff members for their support with various aspects of the project.

% Analia
A.V.S.C. acknowledge financial support from Consejo Nacional de Investigaciones Científicas y Técnicas (CONICET), Agencia I+D+i (PICT 2019–03299) and Universidad Nacional de La Plata (Argentina). A.V.S.C. also thanks FAPESP for the support grant 2025/05085-1.
% Ricardo Demarco
R.D. gratefully acknowledges support by the ANID BASAL project FB210003.
% Liana
L.L. acknowledges the financial support from FAPESP (grants 2024/19951-0 and 2019/26492-3)
% Paulo Lopes
P.A.A.L. thanks the support from CNPq, grants 310260/2025-6 and 404160/2025-5, and FAPERJ, grant E-26/200.545/2023.
% Amanda Lopes
A.R.L. acknowledges the financial support from FAPESP (grant 2025/09544-0) and from CONICET.
% Rodrigo
R.F.H. acknowledges the financial support from CONICET, Agencia I+D+i (PICT 2019–03299) and Universidad Nacional de La Plata (Argentina). 
% Natanael
N.M.C. thanks the support from CAPES (grant 88887.133104/2025-00).
% PAnda
S.P. is supported by the International Gemini Observatory, a program of NSF NOIRLab, which is managed by the Association of Universities for Research in Astronomy (AURA) under a cooperative agreement with the U.S. National Science Foundation, on behalf of the Gemini partnership of Argentina, Brazil, Canada, Chile, the Republic of Korea, and the United States of America. 
% Vinicius Placco
The work of V.M.P. is supported by NOIRLab, which is managed by the Association of Universities for Research in Astronomy (AURA) under a cooperative agreement with the U.S. National Science Foundation. 
% Laerte
L.S.J. acknowledges the support from CNPq (308994/2021-3) and FAPESP (2011/51680-6).
% Pedro Humire
P.K.H. gratefully acknowledges FAPESP for the support grant 2023/14272-4.
% Raimundo Lopes
R.L.O. was partially supportedby CNPq (PQ-315632/2023-2 and 445047/2024-785 0) and FAPESP (2025/01204-6).
% Marcelo Borges
M.B.F. acknowledges financial support from CNPq (grant No. 307711/2022-6).
% Maiara
M.S.C. acknowledges funding from FAPESP grant 2025/12629-8.
% André Santos
A. S. acknowledges the financial support from CNPq (402577/2022-1).
% Lia Dobrawa
L.D. acknowledges the support from the funding agency FAPESP (grant \#2024/03575-9 and \#2025/11378-1).
% Debasish
D.H. acknowledges support from ANID, Chile through grant No. 21232262, which supported Ph.D. studies and a research visit to NOIRLab, Az. 
% Carlos Eduardo
C.E.F.L. acknowledges support from the ANID/FONDECYT Regular grant 1231637, DIUDA project 88231R11, the LSST Discovery Alliance grant, and the GEMINI/ANID grant 32240028.
% Erik Ghuron
E.G. was supported by CAPES (88887.114673/2025-00) 
% Lilianne
L. N. acknowledges FAPESP grant 2024/07281-0
% Gissel
G.M. gratefully acknowledges FAPESP for the support grant 2024/10923-3 and 2025/14602-0
% Barbara
B.C.P. acknowledges the financial support of DIDULS/ULS through the funding ADI2553855.
% Ciria
C. L-D. acknowledges a grant from the Agencia Nacional de Investigación y Desarrollo (ANID) through Fondecyt project 3250511.
% André
A.L.F. acknowledges support from CAPES (grant 88882.332925/2019-01) and FAPESP (grants 2023/06539-0 and 2025/10749-6).
% Raquel
R. R. V. acknowledges the support from CAPES (grant 88887.821818/2023-00) and FAPESP (grants 2023/05003-0 and 2024/16592-9).
% Augusto Damineli
A.D. thanks to FAPESP for the grant 2024/16086-6.
% Daniela Olave
D.E.O-R. acknowledges financial support from ANID Fondo ALMA 2025 project 31250033. D.E.O-R. also thanks ANID InESGénero project INGE210025.
% Felipe
F.A.-F. acknowledges FAPESP grant 2024/00822-5 and acknowledges support from the Brazilian Ministry of Science, Technology and Innovation (MCTI) and the Brazilian Space Agency (AEB), which supported the present work under the PO 20VB.0009.

The author acknowledges the usage of LLM models to improve the grammar, clarity, flow, and consistency in terminology and notation throughout this paper. 

\end{acknowledgments}

\bibliography{sample701}{}
\bibliographystyle{aasjournalv7}

%% This command is needed to show the entire author+affiliation list when
%% the collaboration and author truncation commands are used.  It has to
%% go at the end of the manuscript.
%\allauthors

%% Include this line if you are using the \added, \replaced, \deleted
%% commands to see a summary list of all changes at the end of the article.
%\listofchanges

\end{document}